\documentclass[twocolumn]{aastex631}

\usepackage{newtxtext,newtxmath, diagbox, mathtools}
\usepackage{mathrsfs}
\usepackage{enumitem}

\newcommand{\Bb}{\mathcal{B}}
\newcommand{\Dd}{\mathcal{D}}

\newcommand{\Pp}{\mathscr{P}}

\newcommand{\bea}{\begin{eqnarray}}
\newcommand{\eea}{\end{eqnarray}}

\begin{document}

\title{{UPdec-Webb: A Dataset for Coaddition of JWST NIRCam Images}}

\correspondingauthor{Huanyuan, Shan}
\email{hyshan@shao.ac.cn}

\author[0000-0002-3779-9069]{Lei, Wang}
\affiliation{Purple Mountain Observatory \& Key Laboratory of Radio Astronomy, Chinese Academy of Sciences, 10 Yuanhua Road, Qixia District, Nanjing 210023, China}
\email{leiwang@pmo.ac.cn}

\author[0000-0001-8534-837X]{Huanyuan,Shan}
\affiliation{Shanghai Astronomical Observatory, Chinese Academy of Sciences, 80 Nandan Road, Shanghai 200030, China}
\affiliation{Key Laboratory of Radio Astronomy and Technology, Chinese Academy of Sciences, \\
A20 Datun Road, Chaoyang District, Beijing, 100101, China }
\affiliation{University of Chinese Academy of Sciences, Beijing 100049, China}

\author[0000-0001-6495-1890]{Lin, Nie}
\affiliation{Department of Information Engineering, Wuhan Institute of City, Wuhan, Hubei 430083, China}

\author[0000-0003-0202-0534]{Cheng, Cheng}
\affiliation{Chinese Academy of Sciences South America Center for Astronomy,\\ National Astronomical Observatories, CAS, Beijing 100101, China}
\affiliation{CAS Key Laboratory of Optical Astronomy, National Astronomical Observatories, \\
Chinese Academy of Sciences, Beijing 100101, China}

\author[0000-0001-6763-5869]{Fang-Ting Yuan}
\affiliation{Shanghai Astronomical Observatory, Chinese Academy of Sciences, 80 Nandan Road, Shanghai 200030, China}

\author{Qifan, Cui}
\affiliation{Key Laboratory of Space and  Technology, National Astronomical Observatories, \\
Chinese Academy of Sciences, Beijing 100101, China}

\author{Guoliang, Li}
\affiliation{Purple Mountain Observatory \& Key Laboratory of Radio Astronomy, Chinese Academy of Sciences, 10 Yuanhua Road, Qixia District, Nanjing 210023, China}

\author{Yushan, Xie}
\affiliation{Shanghai Astronomical Observatory, Chinese Academy of Sciences, 80 Nandan Road, Shanghai 200030, China}

\author{Dezi, Liu}
\affiliation{South-Western Institute for Astronomy Research, Yunnan University, Kunming, Yunnan, 650500, China}
\affiliation{University of Chinese Academy of Sciences, Beijing 100049, China}

\author{Yao, Liu}
\affiliation{Purple Mountain Observatory \& Key Laboratory of Radio Astronomy, Chinese Academy of Sciences, 10 Yuanhua Road, Qixia District, Nanjing 210023, China}

\author{Min, Fang}
\affiliation{Purple Mountain Observatory \& Key Laboratory of Radio Astronomy, Chinese Academy of Sciences, 10 Yuanhua Road, Qixia District, Nanjing 210023, China}

\author{Nan, Li}
\affiliation{National Astronomical Observatories, Chinese Academy of Sciences, Beijing 100871, China}

\author[0000-0001-6623-0931]{Peng, Jia}
\affiliation{College of Electronic Information and Optical Engineering, Taiyuan 030024, China}
\affiliation{Peng Cheng Lab, Shenzhen 518066, China}
\affiliation{Department of Physics, Durham University, Durham DH1 3LE, UK}

\author{Ran, Li}
\affiliation{National Astronomical Observatories, Chinese Academy of Sciences, Beijing 100871, China}

\author{Fengshan, Liu}
\affiliation{National Astronomical Observatories, Chinese Academy of Sciences, Beijing 100871, China}

\author{Yiping, Shu}
\affiliation{Purple Mountain Observatory \& Key Laboratory of Radio Astronomy, Chinese Academy of Sciences, 10 Yuanhua Road, Qixia District, Nanjing 210023, China}

\author{Chang, Jiang}
\affiliation{Purple Mountain Observatory \& Key Laboratory of Radio Astronomy, Chinese Academy of Sciences, 10 Yuanhua Road, Qixia District, Nanjing 210023, China}

\author{Cheng-Liang, Wei}
\affiliation{Purple Mountain Observatory \& Key Laboratory of Radio Astronomy, Chinese Academy of Sciences, 10 Yuanhua Road, Qixia District, Nanjing 210023, China}

\author{Han, Qu}
\affiliation{Purple Mountain Observatory \& Key Laboratory of Radio Astronomy, Chinese Academy of Sciences, 10 Yuanhua Road, Qixia District, Nanjing 210023, China}

\author{Wen-Wen, Zheng}
\affiliation{Purple Mountain Observatory \& Key Laboratory of Radio Astronomy, Chinese Academy of Sciences, 10 Yuanhua Road, Qixia District, Nanjing 210023, China}

\author[0009-0006-8959-9161]{Li-Yan Zhu}
\affiliation{Zhejiang Provincial Key Laboratory of Quantum Precision Measurement, College of Science, Institute for Frontiers and Interdisciplinary Sciences,\\ Zhejiang University of Technology, 288 Liuhe Road, Hangzhou 310023, China}

\author{Xi, Kang}
\affiliation{Zhejiang University-Purple Mountain Observatory Joint Research Center for Astronomy, Zhejiang University, Hangzhou 327, China}

\begin{abstract}

We present the application of the image coaddition algorithm, Up-sampling and PSF Deconvolution Coaddition (UPDC), for stacking multiple exposure images captured by the James Webb Space Telescope (JWST) Near-Infrared Camera (NIRCam). By addressing the point spread function (PSF) effect, UPDC provides visually enhanced and sharper images. Furthermore, the anti-aliasing and super-resolution capabilities of UPDC make it easier to deblend sources overlapped on images, yielding a higher accuracy of aperture photometry. We apply this algorithm to the SMACS J0723 imaging data. Comparative analysis with the Drizzle algorithm demonstrates significant improvements in detecting faint sources, achieving accurate photometry, and effectively deblending (super-resolution) closely packed sources. {As a result, we have newly detected a pair of close binary stars that were previously unresolvable in the original exposures or the Drizzled image.} These improvements significantly benefit various scientific projects conducted by JWST. The resulting dataset, named "UPdec-Webb", can be accessible through the official website of the Chinese Virtual Observatory (ChinaVO).

\end{abstract}

\keywords{gravitational lensing: weak--methods: data analysis--instrumentation: James Webb Space Telescope--galaxies: clusters: SMACS J0723 }

\section{Introduction} \label{sec:intro}

Image coaddition stands as a critical task in astronomy, where multiple single-exposure images are aligned and stacked to generate coadded images. This technique substantially improves the analysis of astronomical images of stars and galaxies by enhancing the signal-to-noise ratio (SNR) and reducing noise.

The James Webb Space Telescope (JWST) stands as a pioneering telescope designed to unveil the mysteries of the Universe in the infrared regime. It is equipped with advanced scientific instruments including a Near-Infrared Camera (NIRCam), Near-Infrared Spectrograph (NIRISS), and Mid-infrared Instrument (MIRI), which can revolutionize our understanding of the early Universe, the assembly of galaxies, the planetary system and the origins of life. Webb's primary imager NIRCam's detectors undersample the JWST PSF, which can be particularly severe in the short wavelength (SW) channel, up to a factor of $\sim 2$ with F070W. 

The process of stacking multiple exposures obtained by JWST-NIRCam yields several benefits: 
\begin{enumerate}[label=(\roman*).]

\item{{Increasing the SNR: JWST instruments are highly sensitive to faint infrared emissions from distant astronomical objects. Some scientific goals require detecting very weak signals, necessitating longer exposure times. Deep field observations with JWST involve gathering data over an extended period to reveal the faintest and most distant objects. Image stacking combines data from multiple exposures to improve the SNR, providing prolonged effective exposure times essential for these investigations\citep{Morishita_2023,10.1093/mnras/stac3144}.}}

\item{Reducing Artifacts: Coaddition helps mitigate artifacts and imperfections present in individual under-sampled images. By combining multiple images with different orientations or noise patterns, certain types of artifacts, like cosmic rays, can be cancelled out or reduced, leading to a cleaner final image.}

\item{Enhancing Resolution: JWST's potential for high-resolution imaging is significant. By aligning and co-adding multiple under-sampled images, we can attain spatial resolutions beyond the capability of single exposures. This is essential for resolving detailed structures of distant objects, such as resolving the host galaxies of quasars and the substructures of high-redshift galaxies.}

\end{enumerate}

Numerous coaddition techniques have been developed to enhance image quality from multiple exposures, including Shift-and-Add \citep{Bates+1980,Farsiu+2004a}, Drizzle \citep{Fruchter+2002}, Super-Drizzle \citep{Takeda+2006}, IMCOM \citep{Rowe+2011}, iDrizzle \citep{Fruchter+2011}, SPRITE \citep{Mboula+2015}, fiDrizzle \citep{Wang_2017}, the iterative back-projection \citep{Irani+1993,Symons+2021} and the UPDC \citep{Wang_2022}. Among these methods, Drizzle is a standard and widely adopted approach for combining images captured by the Hubble Space Telescope (HST) and JWST. The UPDC takes the anti-aliasing of under-sampled images and PSF deconvolution into account, similar to a combination of iDrizzle\citep{Fruchter+2011} and Richardson-Lucy methods\citep{1974AJ.....79..745L}. The UPDC approach is designed to tackle key challenges in the coaddition process, particularly focusing on image up-sampling and PSF deconvolution.

The UPDC module was originally developed for data processing of the Multi-channel Imager for the Chinese Survey Space Telescope (CSST-MCI). In this paper, we utilize the UPDC module to combine multiple exposures obtained with JWST NIRCam. With this method, we can coadd the under-sampled exposures to produce a super-resolution image with a higher SNR. This will enhance the ability to detect faint objects at higher redshifts. The UPDC algorithm is used to process multiple exposures from Webb's first deep field, the SMACS-J0723.3-7327 (hereafter referred to as SMACSJ0723) in this paper. For each detector, 9 exposures are taken with approximately 800 seconds for each exposure. This results in 36 frames for the SW and 9 frames for the LW in each of the filters. As part of the Observatory's Early Release Observations \citep{2022ApJ...936L..14P}, several instruments on the JWST observed the cluster on June 7, 2022. The application of UPDC on these data presents an exciting opportunity to showcase its potential for processing JWST NIRCam images.

The paper is organized as follows. In Section \ref{sec:data_reduction} we introduce the data sets and perform the data reduction. Section \ref{sec:flowchat} presents a brief introduction of the flowchart of the UPDC algorithm. In Section \ref{sec:results}, we visually and quantitatively compare the results from Drizzle/Starck-Pantin\citep{Starck+2002} and UPDC on the JWST image reconstruction. Section \ref{sec:summ} summarizes our results. Appendices \ref{errors} and \ref{weights} contain a brief analysis of the uncertainties and weights in UPDC, respectively.

\section{data sets and reduction}
\label{sec:data_reduction}

\begin{figure*}
\centering
\includegraphics[width=1\textwidth,clip,angle=0]{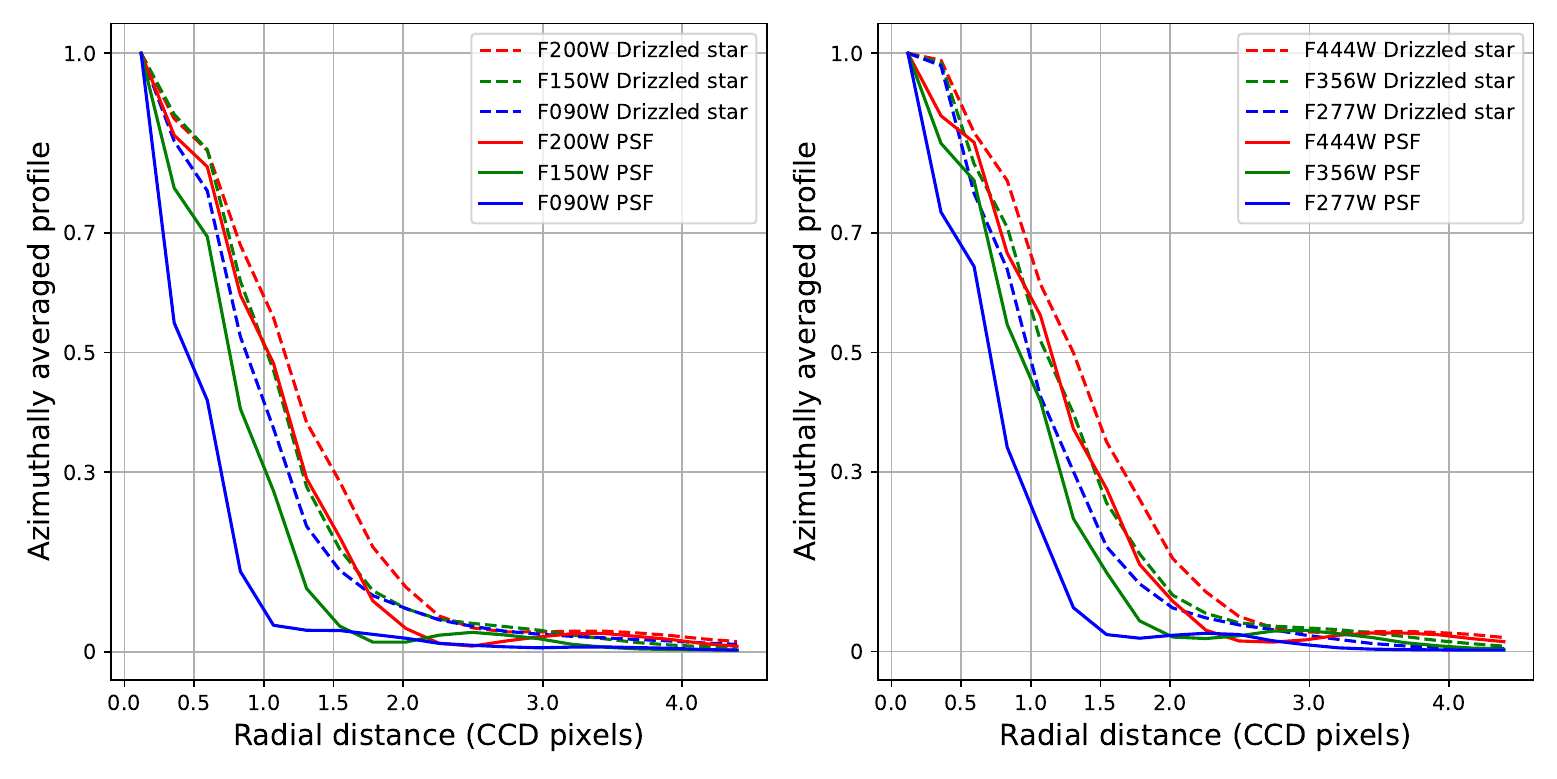}
\caption{{JWST sampling for the six bands. The Nyquist sampling has an FWHM of 2 CCD pixels, namely a radius of 1 CCD pixel. Below this value is under-sampling. The solid lines come from the HybPSF, the dashed from a drizzled star at $R.A.=110^\circ.865243, DEC=-73^\circ.454731$. The left panel represents the SW band (0\arcsec.031/pixel), while the right one for the LW (0\arcsec.063/pixel). All images are sampled doubly compared to their detectors.}}
\label{Nyquistsampling}
\end{figure*}

In this paper, we use the JWST-NIRCam Imaging data of SMACSJ0723 to validate our data reduction pipeline. The NIRCam of the JWST captures images across two channels: the SW channel spanning 0.6 to 2.3 $\mu$m with a resolution of 0\arcsec.031/pixel, and the long wave (LW) channel spanning 2.4 to 5.0 $\mu$m with a resolution of 0\arcsec.063/pixel. Both channels cover a field of view of 9.7 arcmin$^2$, and are viewed simultaneously through the use of a dichroic. JWST will be ideally (Nyquist) sampled at 2 $\mu$m for SW (4 $\mu$m for LW), with two pixels across the FWHM, but will be significantly under-sampled at shorter wavelengths – a situation familiar to many users of HST instruments (see the technical report for JWST PSF)\footnote{\url{https://www.stsci.edu/files/live/sites/www/files/home/jwst/documentation/technical-documents/_documents/JWST-STScI-001157.pdf}}. Figure \ref{Nyquistsampling} shows the sampling cases for all the six bands. The critical sampling (Nyquist) has an FWHM of 2 pixels of the detector, which corresponds to a radius of 1 pixel. Below this value results in under-sampling. The solid lines are derived from the HybPSF (see section \ref{HybPSF}), while the dashed lines come from a drizzled star at $R.A.=110^\circ.865243, DEC=-73^\circ.454731$. The left panel represents the SW band, and the right represents the LW band. All images are sampled doubly compared to their detectors. For the Nyquist sampled images (the red lines), the star profiles are very similar to their PSF. While the pixelation effect caused by under-sampling results in the observed star having extended outer contours, particularly in the severely under-sampled F090W and F277W bands (see also the technical report for JWST PSF). Therefore, in multi-exposure coaddition, upsampling (sub-pixel resampling) is necessary for the under-sampled exposures from F090W, F150W, F277W and F356W bands.

{NIRCam comprises two modules (A and B) separated by a gap of approximately 44\arcsec. Each detector contains $2048\times2048$ pixels, with only the inner $2040\times2040$ pixels used for scientific observations, while the outer 4-pixel wide border serves as references for calibration purposes. The SW channel has 4 detectors, resulting in twice the resolution compared to the LW channel, which has one detector per module. We have downloaded the six bands exposures (Stage-2 data, with \_cal.fits suffix) from the Mikulski Archive for Space Telescopes (MAST) database (i.e., Stage-2: SMACSJ0723)\footnote{\url{https://mast.stsci.edu/search/ui/\#/jwst/results?resolve=true&instruments=NIRCAM&program_id=2736&useStore=false&search_key=7055aba0ea26b8}}. The data are produced by the data processing (DP) software (SDP\_VER '2023\_4a'), the calibration software (CAL\_VER '1.13.3') \citep{2019ASPC..523..543B} and the calibration reference data system (CRDS, CRDS\_VER '11.17.14', CRDS\_CTX '
jwst\_1223.pmap').}

\subsection{{Image correction: replacement for cosmic rays and bad pixels}}\label{Image_correction_sec}
 {Before further processing, we must mark all regions with abnormal flux, like cosmic rays. In this work, we use an improved statistical algorithm which was originally introduced by \cite{Wang_2024} to mark the abnormal regions and replace them with interpolation values. Interpolation is performed among (good) pixels at the same position (overlapped with the abnormal pixel) on other exposures. Statistically, the replacement does not affect the final result compared to the classic approach of marking and assigning zero weight. Another advantage of this approach is that it avoids flux jumps and simplifies handling weight issues (see Appendix \ref{weights}). The $1/{\rm f}$ noise correction is performed with the script by Chris Willott\footnote{\url{https://github.com/chriswillott/jwst}}.}

\subsection{Intracluster light (ICL) processing}
SMACSJ0723 is a galaxy cluster at z = 0.39 discovered as part of the southern extension of the Massive Cluster Survey, with a mass of $M = 8.39\pm0.33\times 10^{14}M_{\odot}$ \citep{Coe_2019} and significant ICL \citep{Montes_2022}. To estimate and subtract the ICL for SMACSJ0723, we use a three-step process. The first step is to measure the initial background of the corrected images, which are generated by the section \ref{Image_correction_sec} (hereafter referred to as the original image), using the $biweight\_location$ estimator of the $Photutils$(Ver.\ 1.12.0, \cite{larry_bradley_2024_12585239})\footnote{\url{https://photutils.readthedocs.io/en/stable/index.html}} Background2D class. We apply this estimator to sigma-clipped boxes of 60 ${\times}$ 60 pixels in a grid across the image, masking only the bad pixels. Subtracting the initial background from the original image can significantly remove the ICL, but it also over-subtracts it. Therefore, we use it only for source detection and the source detection in this step will be unaffected by the ICL. The second step is to adopt the method of four tiers of source detection from \cite{Bagley2023ApJ...946L..12B} to detect the sources in the original image after subtracting the initial background. We use similar parameters as \cite{Bagley2023ApJ...946L..12B}, except for the detection thresholds of 1.5, 1.5, 1.0 and 0.8 for each tier, respectively. The last step is to mask the detected sources obtained by the second step on the original image and estimate the final background using the same $biweight\_location$ estimator of the $Photutils$ Background2D class in sigma-clipped boxes of 60 ${\times}$ 60 pixels in a grid across the image. After applying a median filter across 5 ${\times}$ 5 neighbouring boxes, we create a low-resolution gridded background model. Finally, the $BkgZoomInterpolator$ algorithm is applied to interpolate the filtered array and create a smooth background model. In this article, the coaddition corrected by ICL is not displayed but will appear in subsequent dataset updates.

\subsection{JWST PSF reconstruction: the HybPSF}\label{HybPSF}
\begin{figure}[ht!]
\centering
\includegraphics[width=.5\textwidth,clip]{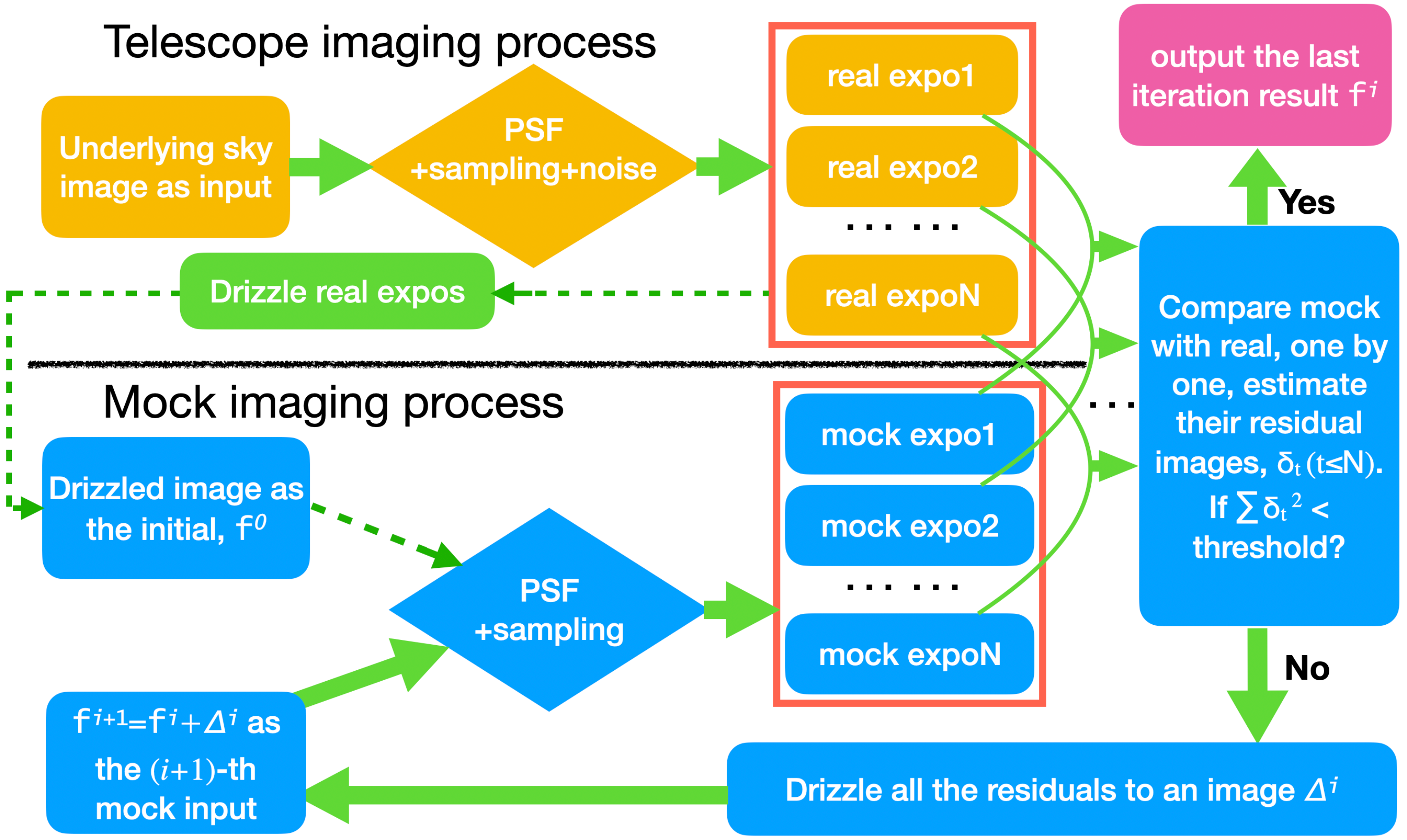}
\caption{A flowchart shows how the UPDC works. The diagram above has two sections, yellow and blue, representing the telescope imaging process and mock imaging process respectively. The dashed green line indicates that this step is only passed through once during initialization. The initial image $f^{(0)}$ (UPDC-iter0) is provided by the JWST pipeline with unshrunk pixels $pixfrac=1$ and double sampling rate ($PSR=0.5$). We convolve the $f^{(0)}$ by the HybPSF and down-sample (or $blot$) the convolved image to mock exposures with the same sampling/dithers as the real exposures (input images). These mocked exposures are then compared one by one with the real exposures. If the comparison results meet the preset threshold, we output the results. If not, we repeat the mock imaging process and use the output to tune the input for the next round of iteration.}
\label{fig:UPDCfc}
\end{figure}

The PSF model serves as a crucial input for the UPDC, aiding in the replication of each exposed image, as illustrated in Figure \ref{fig:UPDCfc}. One notable feature in the JWST NIRCam images is the diffraction spikes originating from stars, which can extend across hundreds of pixels or more. To effectively deconvolve these extended structures using UPDC, the spikes must be adequately represented in the PSF model. Traditionally, constructing the PSF model for each exposure involves isolating unsaturated star images within the image. However, due to the low flux ratio, the spike structures in the PSF are often obscured by noise, making it challenging to accurately model the extended structures using conventional PSF modelling methods.

{WebbPSF ($Ver.\ \rm{1.4.0}$)}, a software tool \citep{2012SPIE.8442E..3DP,2014SPIE.9143E..3XP,2015ascl.soft04007P} that simulates high-quality model PSFs for JWST and Roman based on a combination of observatory design parameters and as-built properties, provides a PSF model with diffraction structures of high fidelity. Despite this, discrepancies may still exist in the inner PSF region between the WebbPSF model and observed star images \citep{2022ApJ...939L..28D,2023ApJ...951...72O,2024AJ....167...58N}. To address this issue, HybPSF was developed to enhance the accuracy of the PSF model for NIRCam images \citep{2024AJ....167...58N}. HybPSF analyzes the residuals between the WebbPSF model and observed star images using PCA \citep{pearson1901liii,2021MNRAS.503.4436N,2021MNRAS.508.3785N}. The reconstructed PSF by HybPSF combines observed data with simulated PSFs from WebbPSF, demonstrating superior performance in terms of PSF shape and profile compared to PCA and WebbPSF. Therefore, we have utilized HybPSF to construct the PSF model for each exposure in this study. It is noteworthy to mention that, akin to the UPDC, HybPSF is also developed as one of the crucial modules in the scientific data processing pipeline of the CSST-MCI.

In the process of constructing the PSF model for UPDC in the mock imaging process, we initially select isolated star images from the download images with $1/{\rm f}$ correction, abnormal pixels masked and zero weight assigned. Then the WebbPSF is employed to generate the input PSF model for HybPSF. {It is worth pointing out that we have checked the shape variations (specifically ellipticity and relative size differences) of the stars in the exposures. The variations between the stars at the centre and edge within each exposure are typically less than $5\%$, while differences across different exposures are about $10\%$. To streamline our processing steps, we employed HybPSF to generate point spread functions (PSFs) centred on each exposure image. These PSFs incorporate an extended spike structure and are sampled onto a 1501$\times$1501 pixel grid with a pixel scale ratio (PSR) of $PSR=0.5$}\footnote{Ratio of input to output pixel scale, see the ResampleStep in the JWST pipeline for details, \url{https://jwst-pipeline.readthedocs.io/en/latest/api/jwst.resample.ResampleStep.html\#jwst.resample.ResampleStep}.}. The size of PSF corresponds to a circle with a radius of 11 arcsecs in the SW band (or 23 arcsecs in the LW band), which almost includes $100\%$ encircled energy. These HybPSFs serve as the input PSFs in the UPDC process.

\section{UPDC flowchart}
\label{sec:flowchat}
As a kind of maximum likelihood estimation (MLE), the UPDC algorithm is proposed and tested by \cite{Wang_2022}, i.e. equation \ref{WLMPHRC}:

\begin{equation}\label{WLMPHRC}
f^{(i+1)} =f^{(i)} \times \frac{1}{L_{\rm E}}  \sum_{k=1}^{L} \Bb^\ast_k \left\{ \Dd^w_k \left\{ \frac{g_k}{ \Dd^h_k\{ \Bb_k\{f^{(i)}\}  \} } \right\} \right\}.
\end{equation}
Where $\frac{1}{L_{\rm E}}\sum_{k=1}^{L} \Bb^\ast_k \left\{ \Dd^w_k \left\{ \cdot  \right\} \right\}$ represents three-step operations in a sequence: up-sampling, convolving PSF and stacking. $f^{(i)}$ is an approximation to the continuous surface brightness of the sky (the underlying image) $f$ in the $i$-th iteration. {According to the theory of MLE\citep{Starck+2002,Fruchter+2011}, in the presence of noise, as the number of iterations increases, the iterates first approach the unknown object and then may potentially move away from it. The iterations can be considered as a regularization parameter. Therefore, there exists an optimal number of iterations at which the SNR reaches its maximum or the total variance is minimized. After the optimal number of iterations, signals in residual will diminish, while amplified noise will dominate.}

In Figure \ref{fig:UPDCfc}, a flowchart briefly shows how the algorithm works. The UPDC mocks the telescope imaging progress, e.g., the PSF convolution and sampling (pixelation), and compares the mocks to the real exposures one by one. The real exposures (or input images) have been corrected with the $1/{\rm f}$ effect and abnormal flux replacement in section \ref{Image_correction_sec}. These effects do not need to be simulated in UPDC. The comparison will give feedback to the input of the next iteration until the result meets our threshold. {Here the initial condition of the iteration is set as the stacking image of the real exposures, which is $Drizzled$ by the ResampleStep in the JWST pipeline with unshrunk pixels $pixfrac=1$ and double sampling rate $PSR=0.5$. In the residual drizzling step, the $Drizzlepac$($Ver.\ 3.7.0$; $pixfrac=1$ and $PSR=0.5$) is used \citep{2010bdrz.conf..382F,2021AAS...23821602H} and all weights are set to be equal (see Appendix \ref{weights}).}

Following \cite{Wang_2022}, we introduce a non-strict positivity constraint $\mathbb{R}^+$ to ensure the smooth running of UPDC:
\begin{equation}\label{cstr}
\Pp_{\mathbb{R}^+}\{ f^{(i+1)} \}=
\begin{cases}
f^{(i+1)},&(f^{(i+1)} \ge 0)\\
f^{(i)},&(f^{(i+1)} < 0)
\end{cases}
\end{equation}
where $\Pp_{\mathbb{R}^+}\{ f^{(i+1)} \}$ is a component-wise projection of $f^{(i+1)}$ onto the set $\mathbb{R}^+$.

\section{Results}
\label{sec:results}
After taking into account the effects mentioned earlier, we use the UPDC to combine several exposures into a double sampling grid ($PSR=0.5$) in each band. This $PSR$ ensures a critical (Nyquist) sampling for F090W and full sampling for F277W bands, as shown in Figure \ref{Nyquistsampling}. Specifically, we combine 36 exposures in each band for the SW channel and 9 exposures in each band for the LW channel. The final output is a result of 10 iterations, each with an image size of $9584\times 9584$ pixel (SW) or $4752\times 4752$ pixel (LW). Here again, the initial condition UPDC-iter0 is from the ResampleStep in the JWST pipeline with unshrunk pixels ($pixfrac=1$) and double sampling rate ($PSR=0.5$).

\begin{figure*}
\centering
\includegraphics[width=1\textwidth,clip]{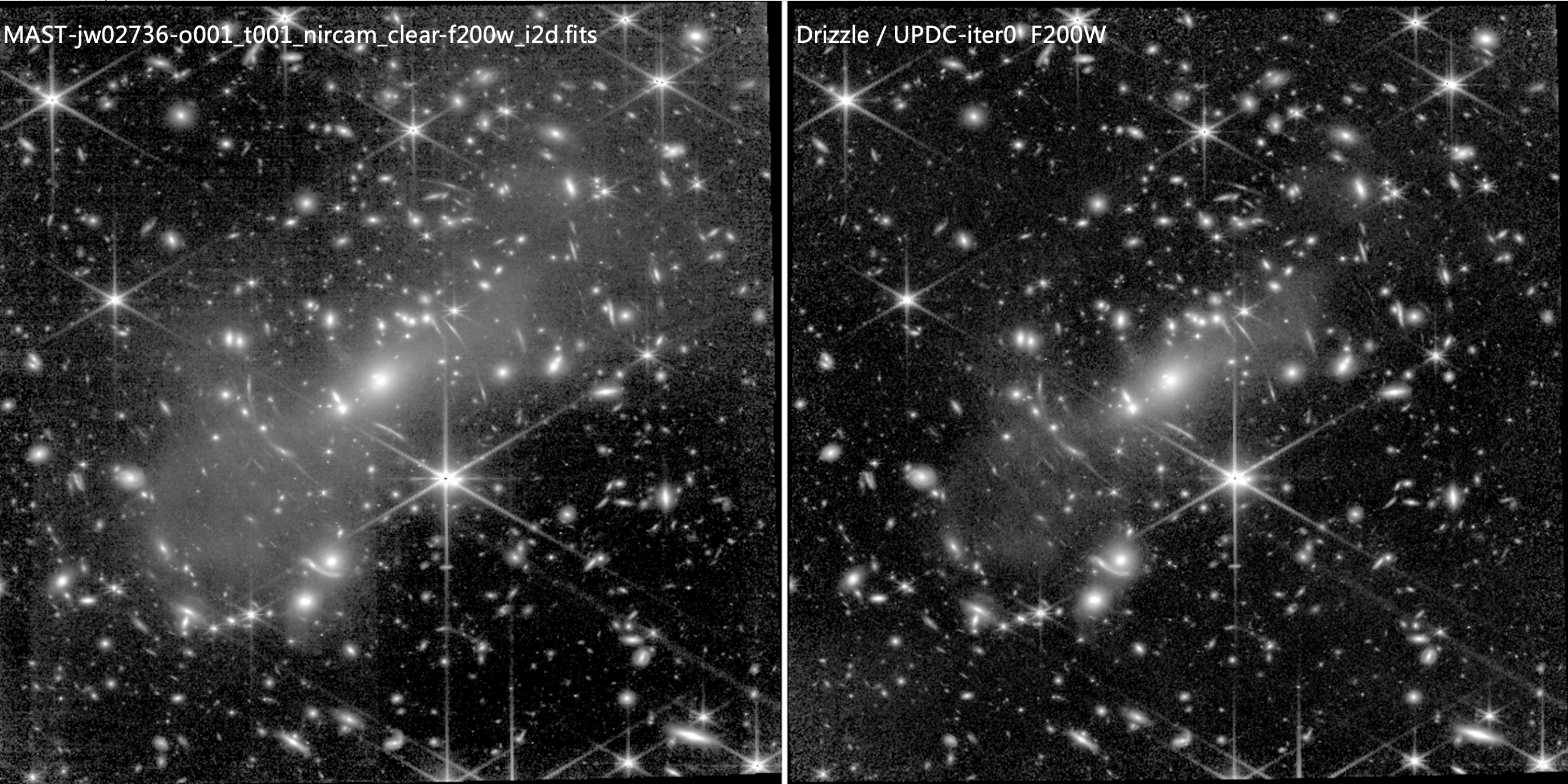}
\caption{{Comparison of the UPDC-iter0 image with the mosaic released on MAST {\it jw02736-o001\_t001\_nircam\_clear-f200w\_i2d.fits}. For comparison purposes, the right panel is output with the same sampling rate ($PSR=1.0$) as the left one. The image on the right panel displays a better background subtraction. It also shows a more continuous distribution of ICL in comparison to the image on the left. The $1/{\rm f}$ effect is visually apparent on the left panel.}}
\label{mastdb-updc}
\end{figure*}

In Figure \ref{mastdb-updc}, we compared the output from the UPDC-iter0 (right panel, which is equivalent to the Drizzle algorithm) to the official released stacking data from the MAST (left panel)\footnote{\dataset[doi:10.17909/7wqf-xd87]{https://doi.org/10.17909/7wqf-xd87}}. For comparison purposes, we output the right panel with the same sample rate ($PSR=1.0$) as the left one. The image on the right panel shows a higher consistency of the output image's background in the coaddition of the four fields of view and a more continuous distribution of ICL than that on the left. The continuity distribution of ICL within the galaxy cluster has been confirmed in the LW channel \citep{Montes_2022}. The $1/{\rm f}$ effect is very noticeable on the left panel. 

\begin{figure*}
\centering
\includegraphics[width=1\textwidth,clip]{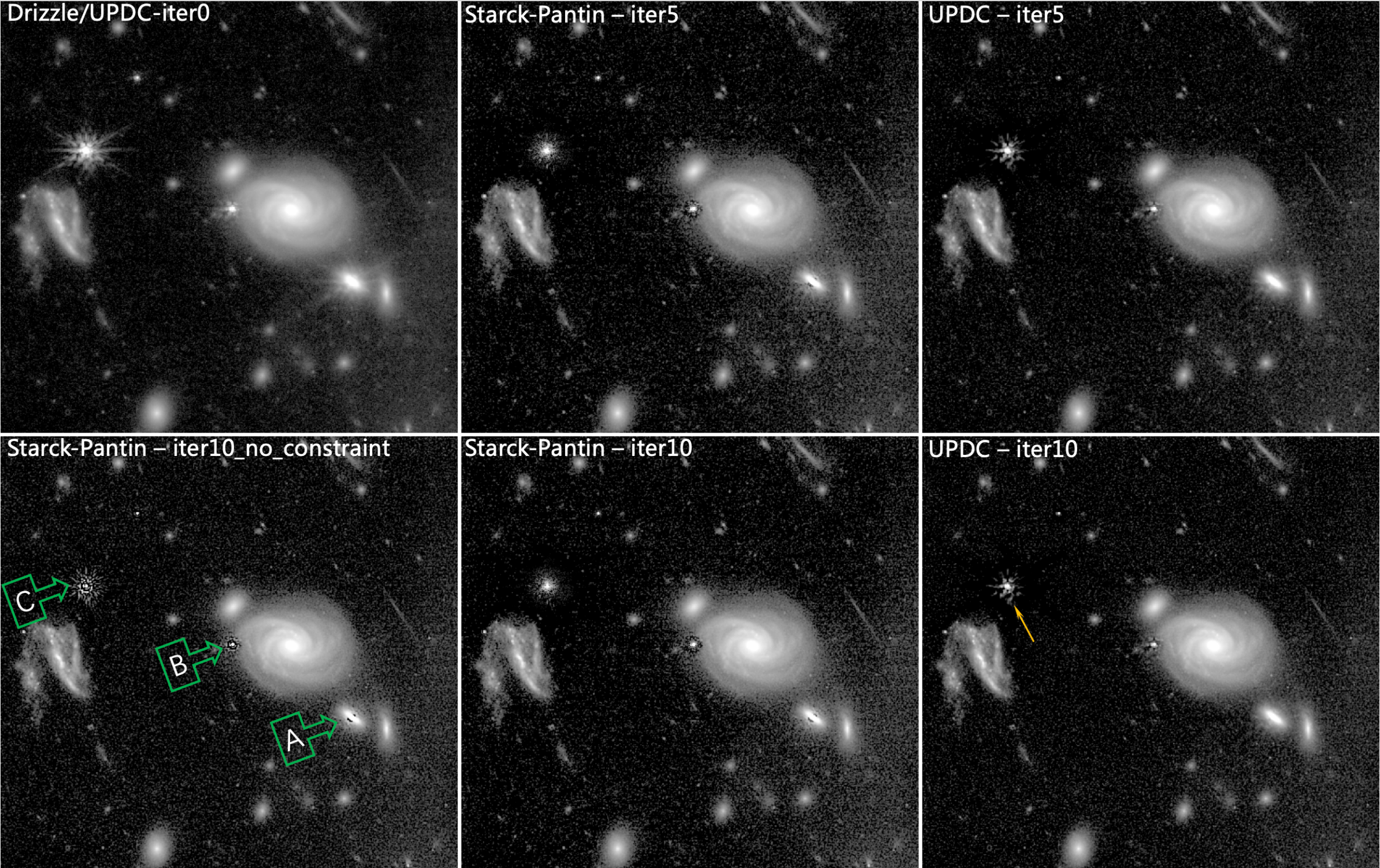}
\caption{{Coaddition comparison between the Starck-Pantin and the UPDC in the F200W band. All images except for the left column have been subjected to a non-strict positivity constraint Eq. \ref{cstr}. It is evident that both methods effectively suppress the diffraction spikes surrounding bright sources; however, the Starck-Pantin method results in increased noise levels. {\bf Tag-A, B, C} illustrates that the Starck-Pantin algorithm induces more pronounced ringing artifacts compared to UPDC, significantly impacting the surface brightness distribution of galaxies, as highlighted by {\bf Tag-A}. In the case of point-source reconstruction, depicted in {\bf Tag-B, C}, the Starck-Pantin approach disrupts the PSF's original structure, substituting it with ringing and noise enhancements. Conversely, the UPDC algorithm preserves the PSF's self-similarity throughout equivalent iterations. As indicated by the yellow arrow, the self-similar PSF allows us to more easily distinguish between the PSF and faint sources blended with it.}}
\label{SPvsUPDC}
\end{figure*}

{We compared the UPDC with the Starck-Pantin algorithm \citep{Starck+2002} for stacking multiple exposures in the F200W band and displayed the results in Figure \ref{SPvsUPDC}. It can be seen that both methods effectively suppress the diffraction spikes around bright sources, but the Starck-Pantin method introduces stronger noise. {\bf Tag-A, B, C} shows that the Starck-Pantin algorithm causes stronger ringing effects compared to the UPDC, also reported in the previous work \citep{Wang_2022}, which even affects the surface brightness distribution of galaxies, as shown by {\bf Tag-A}. In the reconstruction of point sources, as shown in {\bf Tag-B, C}, the Starck-Pantin algorithm destroys the original structure of the PSF, replacing it with ringing and noise, while the UPDC algorithm maintains the self-similarity of the PSF after the same iterations. The self-similarity ensures the predictability of the PSF changes after each iteration, providing a basis for modelling the PSF on the stacked image.}

\begin{figure*}
\centering
\includegraphics[width=1\textwidth,clip]{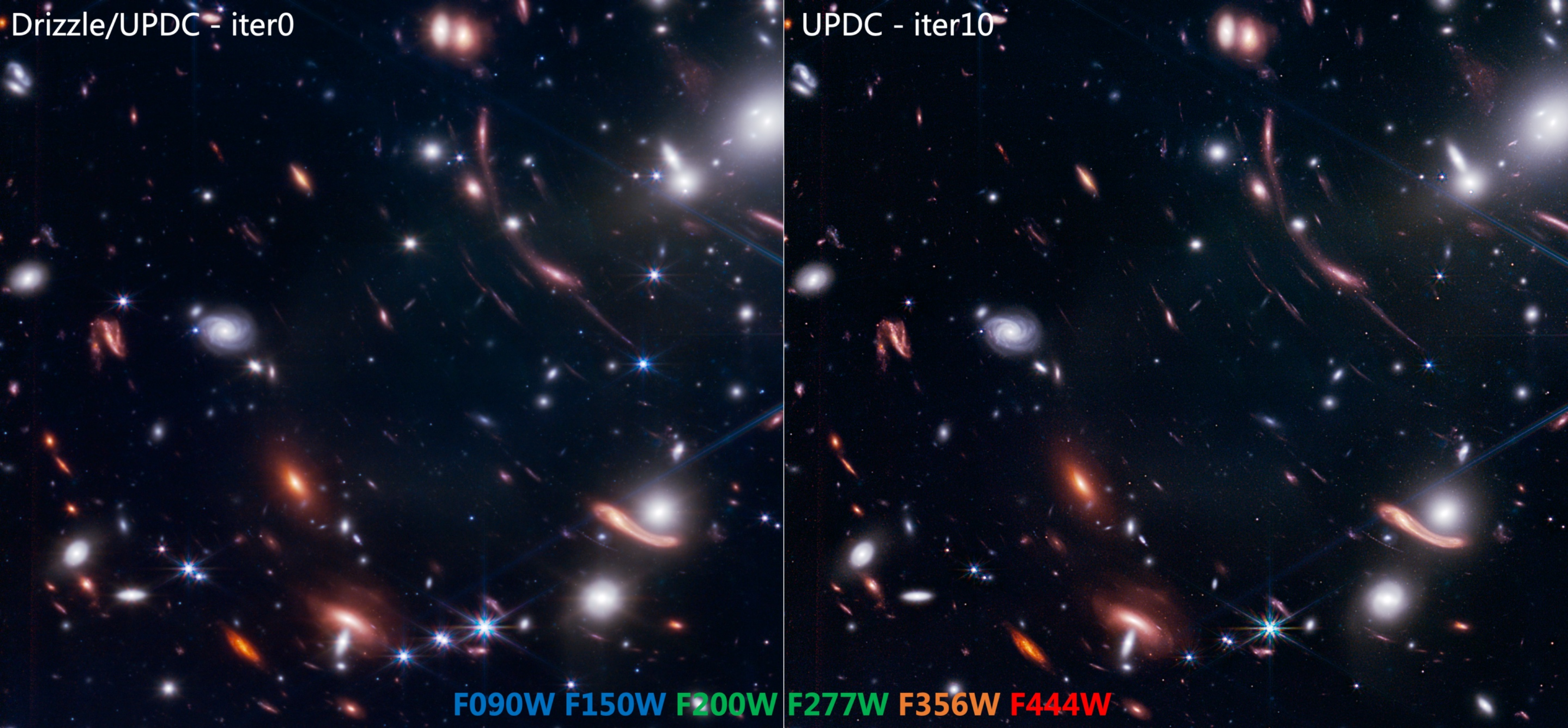}
\caption{A corner of the SMACSJ0723 in RGB image. We mapped grey-scale images to monochromatic colours. Four colour channels are used here (Red: F444W, Orange: F356W, Green: F277W+F200W, Blue: F150W+F090W). {The RGB flux is combined by the following: Blue = F150W + F090W; Green = F200W + F277W + (165/255)$\times$F356W; Red = F356W + F444W.} The left panel is a Drizzle combined image corresponding to the official NASA press release image. The UPDC's result is shown on the right panel with 10 iterations. }
\label{SM0723-1}
\end{figure*}

\begin{figure*}
\centering
\includegraphics[width=1\textwidth,clip,angle=0]{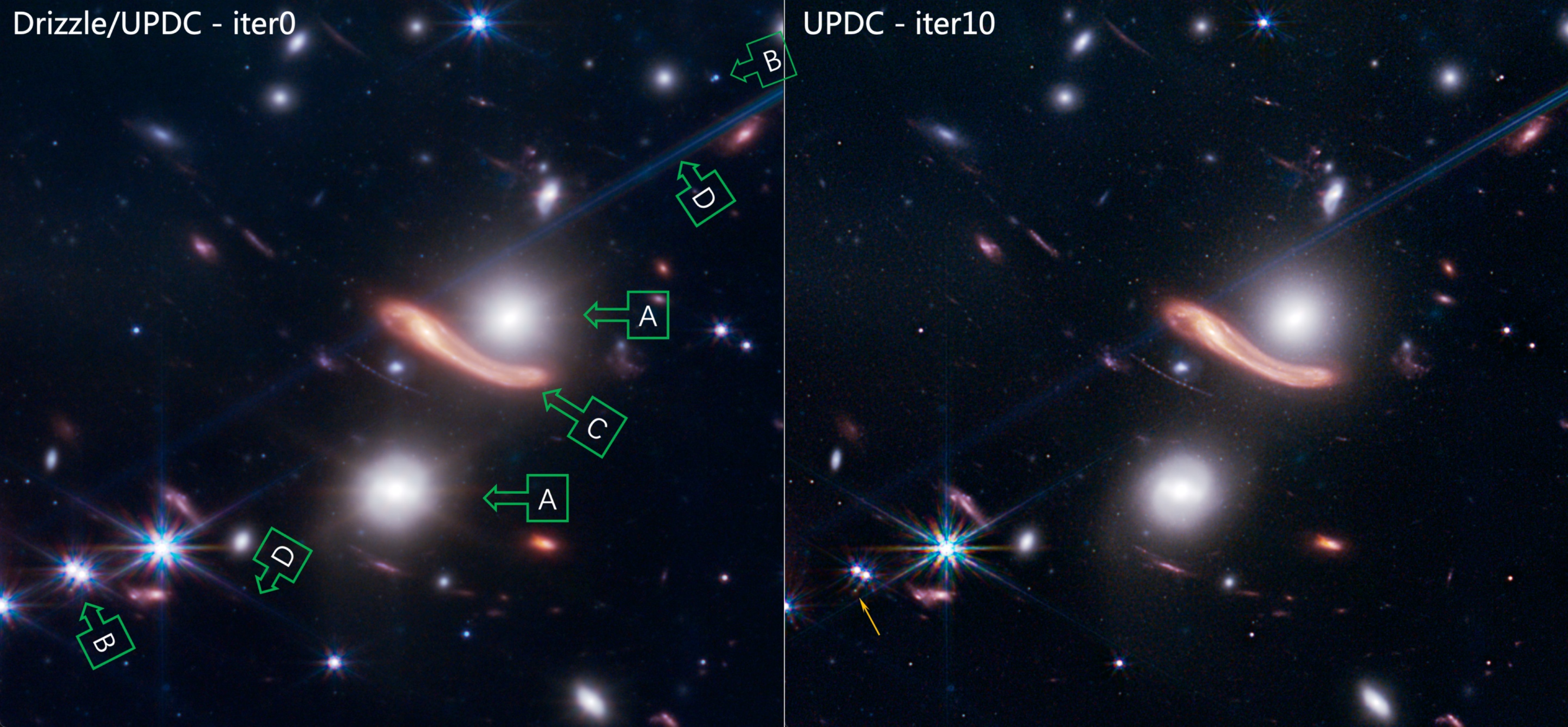}
\caption{A zoom-in RGB image for some details in the SMACSJ0723 field. The JWST diffraction spikes on the elliptical and face-on spiral indicated by {\bf tag-A} have been significantly reduced by UPDC; the two sets of binary star systems marked by {\bf tag-B} have been clearly distinguished; the background galaxy distorted by strong gravitational lensing, as indicated by {\bf tag-C}, has become clearer; as shown by {\bf tag-D}, since the UPDC algorithm is based on a forward modelling Bayesian estimation, the spikes of the central saturated and truncated flux stars cannot be effectively suppressed by PSF deconvolution; finally, the distinguishable point sources marked by the yellow arrow in the right panel are obscured by bright spikes in the left.}
\label{SM0723-2}
\end{figure*}

\begin{figure*}
\centering
\includegraphics[width=1\textwidth,clip,angle=0]{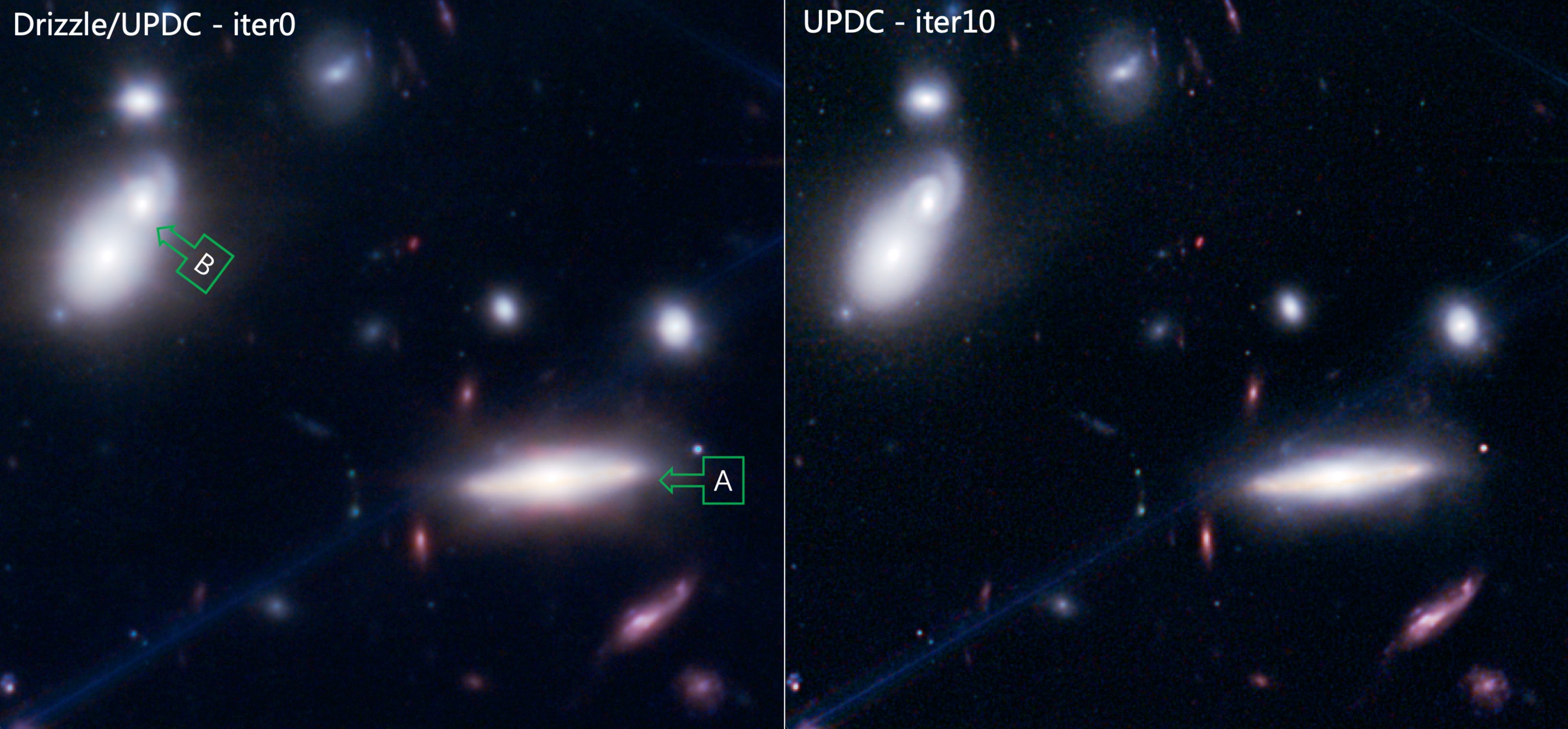}
\caption{Enhancement for the details of galaxies in the SMACSJ0723 field. As shown in {\bf tag-A}, the dust at the edge of the spiral galaxy, which is originally obscured by PSF blurring in the left panel, has become discernible in the right one (the yellow belt traversing the galaxy). Two spiral arms of the spiral galaxy, which are originally blurred in the upper left, have also become very distinguishable in the right panel generated by UPDC, as indicated in {\bf tag-B}.}
\label{SM0723-3}
\end{figure*}
 
\subsection{Enhancement of the image fidelity}

RGB images can be created by combining three LW bands and three SW bands, similar to the images released on the JWST's official website for its first light\footnote{\url{https://webbtelescope.org/contents/media/images/2022/035/01G7DCWB7137MYJ05CSH1Q5Z1Z}}. However, the pixel sizes of the LW and SW band images differ, requiring multiple exposures of all six bands to produce an RGB image on the same output grid, i.e., $4752\times 4752$ pixel. In Figure \ref{SM0723-1}, a corner of SMACSJ0723 is shown, with blue colour contributed by filters F150W and F090W (equally weighted), and green colour represented by filters F277W and F200W (equally weighted). The orange colour (RGB value: 255,165,0) comes from the F356W band and the red colour from the F444W band\footnote{{The RGB flux is combined by the following: Blue = F150W + F090W; Green = F200W + F277W + (165/255)$\times$F356W; Red = F356W + F444W.}}, resulting in an RGB image that matches the official NASA image. The left panel in Figure \ref{SM0723-1} shows a Drizzle combined image consistent with the NASA release, while the right panel displays the output of UPDC after ten iterations of approximation. This colour setting will be used throughout the article for all RGB images. Upon preliminary examination, the image output by UPDC (right image) presents superior quality in comparison to the image output by Drizzle (left image). Notably, the UPDC image exhibits sharpened features, heightened contrast, enriched spatial details of objects, and a more authentic morphology of objects. More details will be shown in the following figures.
 
\begin{figure*}
\centering
\includegraphics[width=1\textwidth,clip,angle=0]{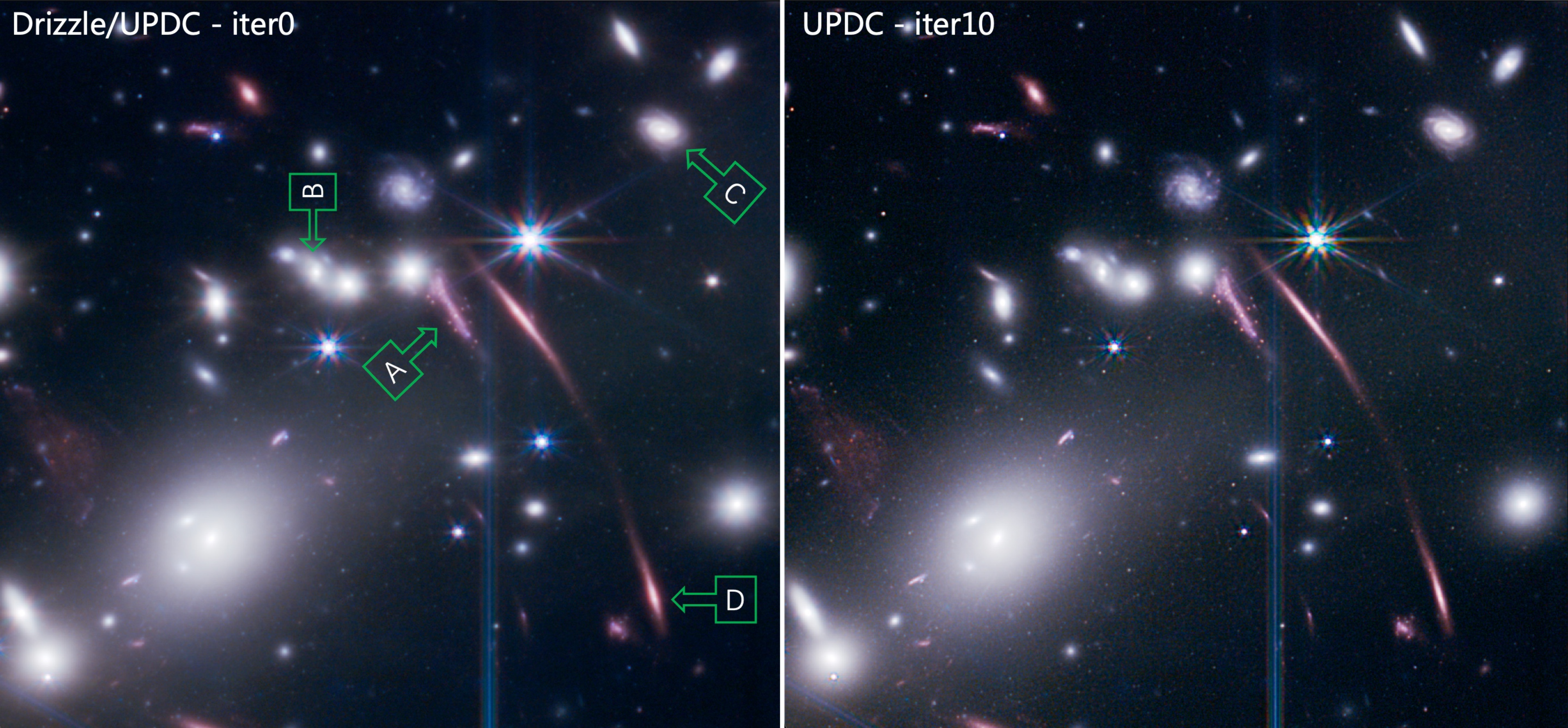}
\caption{Zoomed-in objects near the centre of SMACSJ0723. The UPDC processing reveals clearer globular clusters in the Sparkler Galaxy, as shown in {\bf tag-A}. {\bf Tag-B} and {\bf tag-C} display reconstructed spatial details and additional substructures in galaxies processed with UPDC software. Additionally, the gravitational lensing of a galaxy cluster results in a stretched galaxy forming a giant arc, which becomes slimmer after UPDC processing, as depicted in {\bf tag-D}.}
\label{SM0723-4}
\end{figure*}

\begin{enumerate}[label=(\roman*).]
\item{In Figure \ref{SM0723-2}, the UPDC has significantly reduced the JWST characteristic diffraction spikes, the "snowflake" pattern, indicated by {\bf tag-A}. Additionally, the two sets of binary star systems marked by {\bf tag-B} are now clearly distinguishable. Also, the background galaxy, which was previously distorted by strong gravitational lensing as indicated by {\bf tag-C}, is now much clearer. However, the spikes of the central saturated and truncated flux stars cannot be suppressed by PSF deconvolution since the UPDC algorithm is based on a forward modelling Bayesian estimation, as shown by {\bf tag-D}. Lastly, the distinguishable point sources marked by the yellow arrow in the right panel were previously obscured by bright spikes in the left panel.}

\item{In the left panel of Figure \ref{SM0723-3}, the dust at the edge of the spiral galaxy is obscured by PSF blurring as shown in {\bf tag-A}, but it has become visible in the right panel (the yellow belt traversing the galaxy). Similarly, in the upper left, the two spiral arms of the galaxy are originally blurred (as indicated in {\bf tag-B}) but in the right image produced by UPDC, they have become much more distinguishable.}

\item{ In Figure \ref{SM0723-4}, more objects and richer details can be observed closer to the centre of SMACSJ0723. Take the Sparkler Galaxy's globular clusters, for example, they become increasingly clear, allowing for the discernment of even more of them, as shown in {\bf tag-A}. {\bf Tag-B} features a barred spiral galaxy with an inner bulge and outer disk that has had its spatial details reconstructed after being processed with UPDC. Meanwhile, the spiral galaxy marked by {\bf tag-C} has more substructures. In the background, a galaxy has been severely stretched by the gravitational lensing of the galaxy cluster as shown in {\bf tag-D}, forming a giant arc in which the high-brightness part becomes fat due to PSF blurring. After UPDC processing, the giant arc becomes slimmer than before. }

\begin{figure*}
\centering
\includegraphics[width=1\textwidth,clip,angle=0]{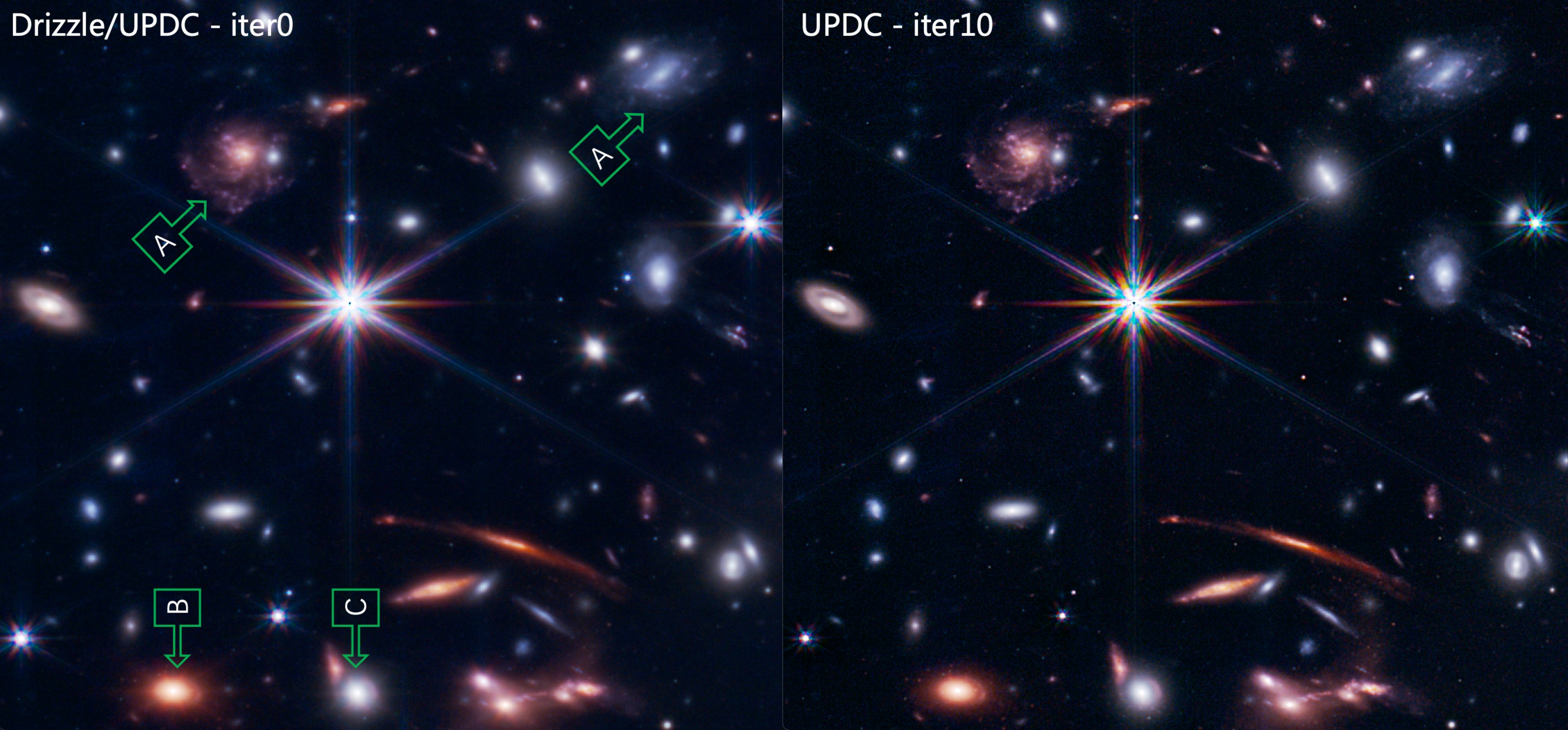}
\caption{Enhancement for spiral galaxies and a strong gravitational lens system. The UPDC reconstructed images reveal richer structural details of two spiral galaxies - red and blue ({\bf tag-A}). The peripheral structure of the red galaxy, hidden by its spikes, is now visible in the output images ({\bf tag-B}). The UPDC also helped detect a candidate for a strong gravitational lens system {(Deng, Shu \& Wang, 2024, in preparing)} by reconstructing the background galaxies that were originally obscured by the spikes of the foreground high-brightness lens galaxy ({\bf tag-C}).}
\label{SM0723-6}
\end{figure*}

\item{In Figure \ref{SM0723-6}, two spiral galaxies, one red and one blue, are identified by {\bf tag-A}. The reconstructed images by UPDC show that these galaxies possess richer substructures. {\bf Tag-B} marks the peripheral structure of the red galaxy which was previously hidden by its spikes, but UPDC has recovered it in the right panel. Furthermore, the UPDC algorithm has enabled the detection of a candidate for a strong gravitational lens system, by reconstructing the background galaxies that were obscured by the spikes of the foreground high-brightness lens galaxy, as illustrated in {\bf tag-C}. Lens modelling and relative measurements are currently being conducted on this strong gravitational lens system by our collaborators.}

\item{Figure \ref{SM0723-8} shows how UPDC enhances the resolving power of objects that overlap each other. The red galaxy, marked by {\bf tag-A}, displays more distinct point-like sources after UPDC processing. The white object located at the lower right of the bulge may be a foreground/background galaxy. The UPDC effectively deblurs the spiral galaxy marked by {\bf tag-B}, separating it from the foreground stars and the galaxy to its right. Our algorithm not only suppresses the central spikes of the elliptical galaxy marked by {\bf tag-C}, but also well separates a galaxy from its upper left. The star marked by {\bf tag-D}, where the part of the green color (F277W and F200W bands) has reached saturation and the central flux is truncated, results in the green part of the PSF not being deconvolved. Consequently, the green part of the spikes is more extended than other colors.}

\begin{figure*}
\centering
\includegraphics[width=1\textwidth,clip,angle=0]{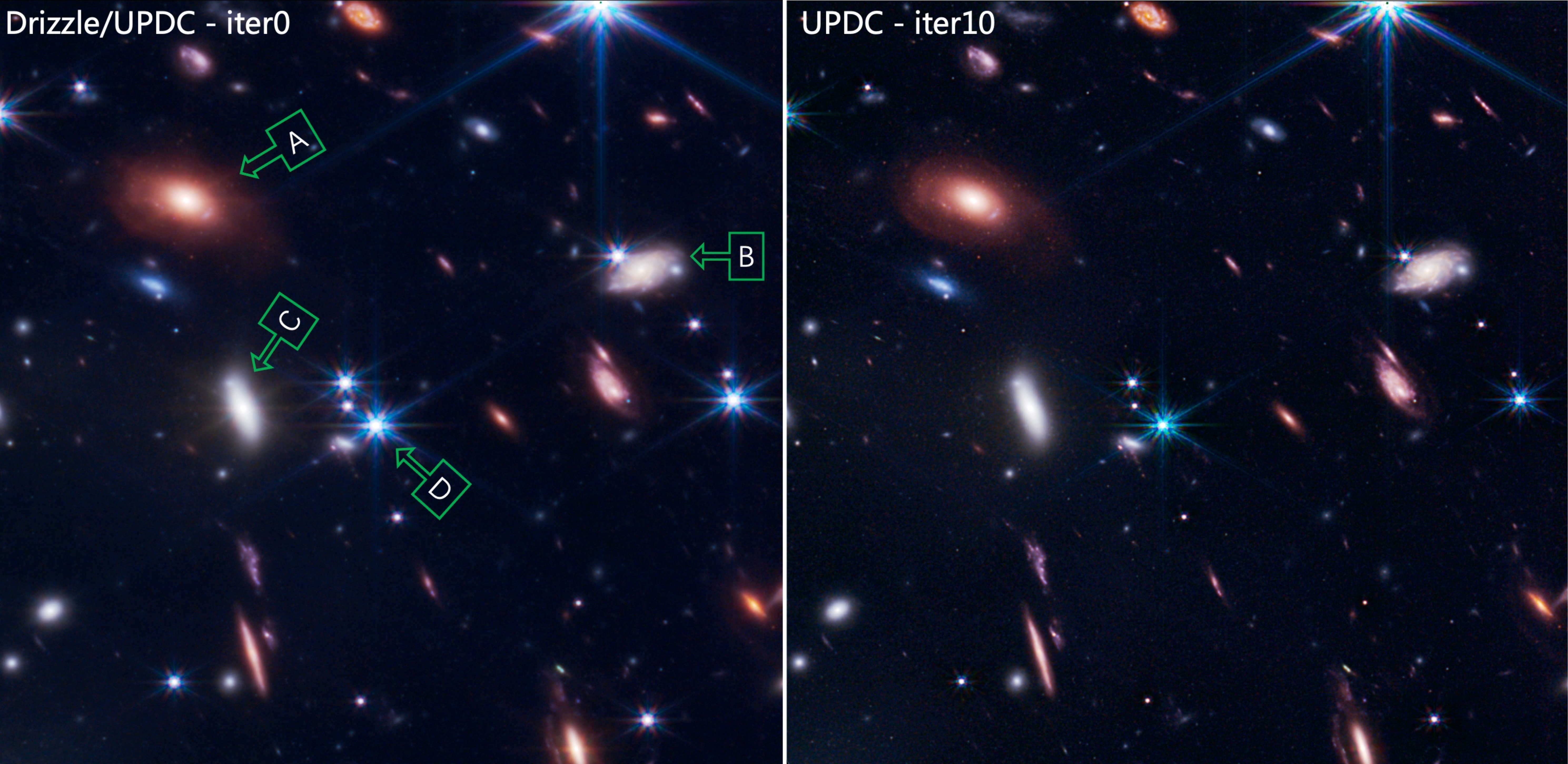}
\caption{Deblending for overlapped objects. The red galaxy labelled by {\bf tag-A}, exhibits clearer point sources following UPDC processing, which are generally believed to be globular clusters, such as the famous Sparkler Galaxy in Figure \ref{SM0723-4}. There is a possibility that the white object situated at the lower right of the bulge is a foreground/background galaxy. {\bf tag-B}, the spiral galaxy, is effectively freed from blurring as a result of UPDC processing, allowing it to be distinguished from both the foreground stars and the galaxy to its right. The UPDC algorithm not only suppresses the central spikes of the elliptical galaxy ({\bf tag-C}), but also separates a galaxy from its upper left. {\bf Tag-D}'s star in the UPDC-iter10 has an obvious green colour, due to the central flux missing in F277W and F200W bands. This causes the green component of the PSF to remain uncorrected by UPDC.}
\label{SM0723-8}
\end{figure*}

\begin{figure*}
\centering
\includegraphics[width=1\textwidth,clip,angle=0]{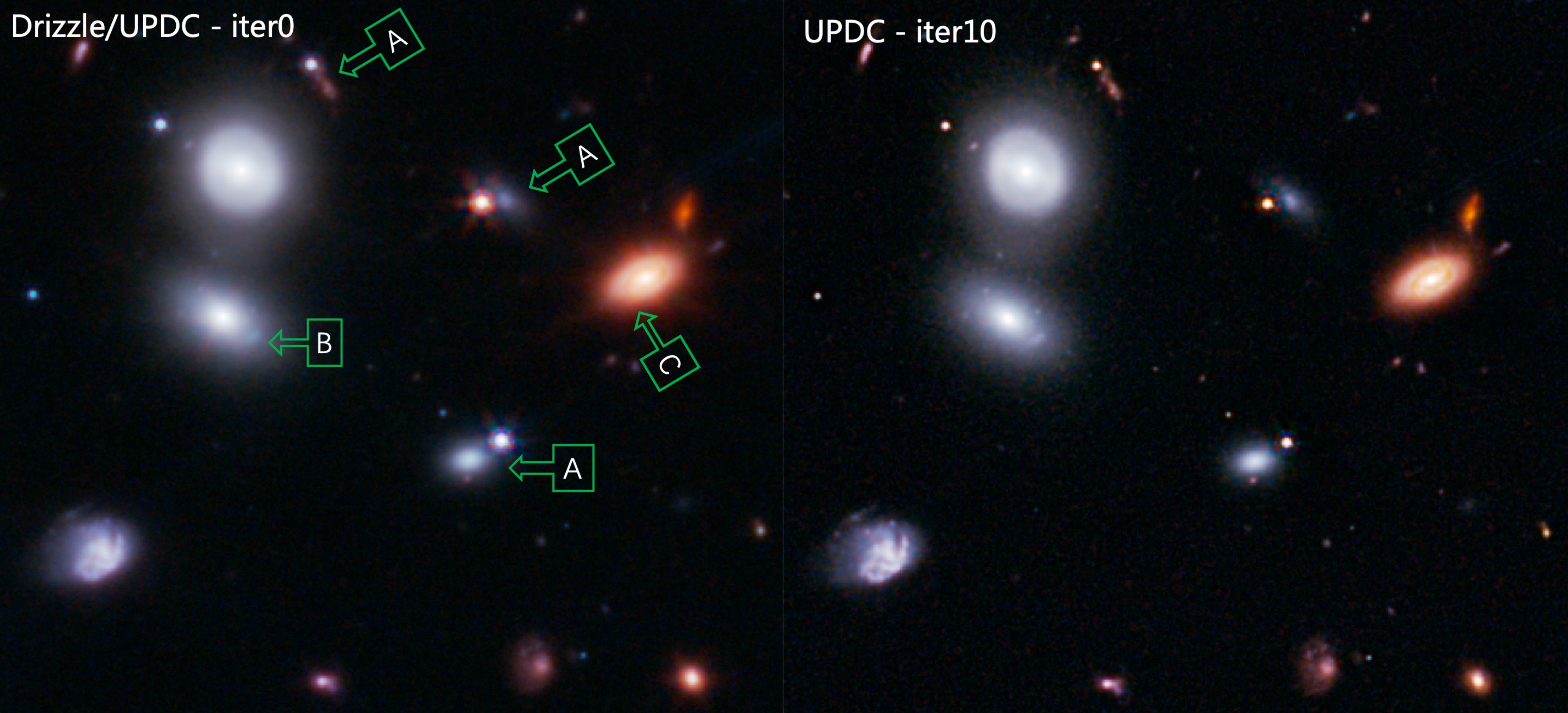}
\caption{Enhancement for objects in the parallel field of SMACSJ0723. Three pairs of objects have been designated as {\bf tag-A} and display differing levels of overlap in their line of sight. However, thanks to the implementation of the UPDC algorithm, each pair of objects is now more distinct, thus enabling an improved photometric analysis. Moreover, the UPDC algorithm has successfully identified a substructure within the galaxy located at {\bf tag-B}. This substructure may either be a foreground or background galaxy or part of the galaxy itself. Originally, the galaxy featured a luminous bulge and surrounding disk, however, following our pipeline processing, it now exhibits a clear two-arm spiral structure, evidenced at {\bf tag-C}.}
\label{PF-1}
\end{figure*}

\item{In this work, we processed the SMACSJ0723 frames captured by the JWST-B module using UPDC, and also handled the exposures taken by the JWST-A module, the parallel field of SMACSJ0723. In Figure \ref{PF-1}, there are three pairs of objects marked as {\bf tag-A} that exhibit varying degrees of overlap along the line of sight. However, after using the UPDC, the objects in each pair are more separated, which allows for better photometric analysis. Additionally, our pipeline has identified a substructure in the galaxy at {\bf tag-B}. This substructure could be a foreground or background galaxy or part of the galaxy itself. Originally, the galaxy had a bright bulge and surrounding disk, but after processing through our pipeline, it now shows a clear two-arm spiral structure, as seen in {\bf tag-C}.}

\end{enumerate}

\begin{figure*}
\centering
\includegraphics[width=1\textwidth,clip,angle=0]{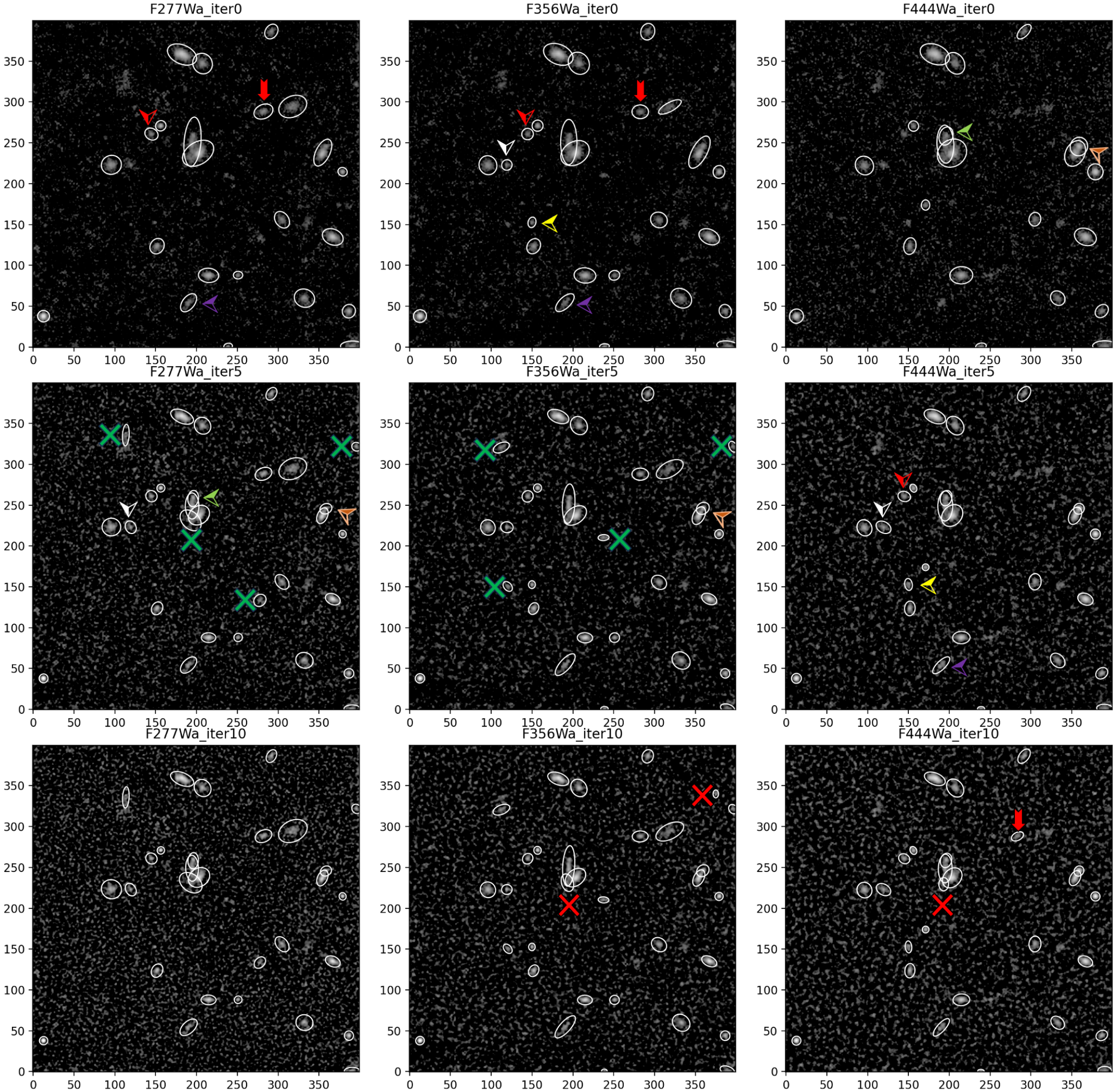}
\caption{UPDC enhances the signal of faint sources. The detected sources are indicated by white ellipticals, i.e. the Kron apertures\citep{Kron_1980}. In the top row of stamps, 8 faint sources marked by coloured arrows are undetected in other bands (Drizzle images, i.e. UPDC-iter0 images) but are identified after UPDC processing (UPDC-iter5 images). This reflects the improved performance of UPDC in detecting faint sources. The panels in the middle row exhibit about 25\% more new detections (16 sources) than those in the upper row. Half of the 16 new detections have been cross-matched in multiple bands, while the remaining half marked with green crosses have not been cross-matched, but exhibit 6$\sigma_{\rm bkg}$ over the background. {In the bottom row of panels, there are minimal new detections in the UPDC-iter10 coadditions compared to the UPDC-iter5 images, with only 4 additional cases in total. The coordinate axis is measured in pixels, with each pixel corresponding to 0\arcsec.031. The centre of the stamp is located at $R.A.=110^\circ.694059, DEC=-73^\circ.475037$.}}
\label{detection}
\end{figure*}

\begin{figure*}
\centering
\includegraphics[width=1\textwidth,clip,angle=0]{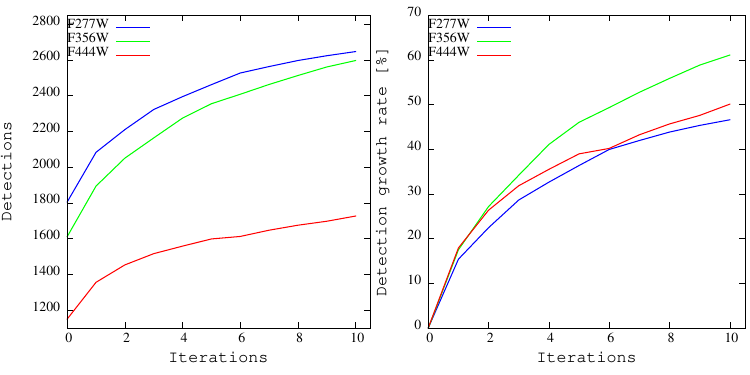}
\caption{The dependency of detection upon UPDC iterations. The left panel displays total detection numbers for three LW bands while the right panel shows corresponding detection growth rates, normalized to the Drizzle image's detection number (UPDC-iter0).}
\label{det_rate}
\end{figure*}

\subsection{Improvement for source detection and photometry}

The UPDC algorithm provides not only sharper images but also significant improvements in detecting faint sources and photometry. In this section, we will discuss the benefits of the UPDC algorithm in detecting faint sources, improving spatial resolution, and enhancing photometry in the exposures coaddition, using the three LW bands of NIRCam: F277W, F356W, and F444W as examples. We have chosen the parallel field of SMACSJ0723 captured by the JWST Module A for comparison, to avoid any interference caused by the ICL or the strong gravitational lensing of galaxy clusters.

\begin{enumerate}[label=(\roman*).]
\item{Figure \ref{detection} shows how the UPDC enhances the signal of faint sources. White ellipticals mark the Kron apertures\citep{Kron_1980} for detected sources. Some of the new detections in the reconstructed images are confirmed by using multi-band source cross-matching. The top row of images shows the Drizzle coaddition (UPDC-iter0) of three LW bands of JWST-NIRCam, where 9 exposures were coadded in each band. Among them, 8 faint sources marked by coloured arrows in the top row are undetected in the Drizzle coaddition images of other bands. However, they are all detected after UPDC processing. The white ellipse represents the source detected by the routine of "image segmentation" ($photutils.segmentation$) in the photometry software $Photutils$. The detection threshold parameters for the two rows of images are set to be the same, reflecting the improvement of our pipeline in detecting faint sources. In the middle row of panels, there are approximately 25\% more new detections, with a total of 16 sources. Out of these 16 new detections, half of them have been identified in multiple bands, and another half of them have not yet been identified i.e., marked with green crosses. {In the bottom row of panels, there are minimal new detections in the UPDC-iter10 coadditions compared to the UPDC-iter5 images, with only 4 additional cases in total (one of them is cross-matched in three LW bands). However, the noise in the UPDC-iter10 coadditions has significantly increased. The detection threshold parameter for this field is set to be 6.0 times the background ($\sigma_{\rm bkg}$) in the $detect\_sources$ function of $Photutils$. This threshold has remained unchanged in the following tests.}}

\begin{figure*}
\centering
\includegraphics[width=1\textwidth,clip,angle=0]{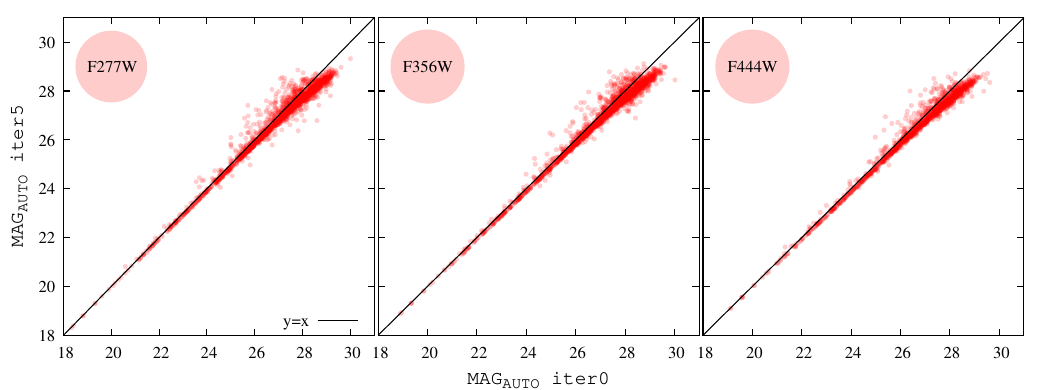}
\caption{The photometry of cross-matched sources in the parallel field of SMACSJ0723 is compared between UPDC-iter0 and UPDC-iter5. Three panels are for three LW bands: F277W, F356W and F444W.}
\label{fig_mag_comp}
\end{figure*}

\begin{figure*}
\centering
\includegraphics[width=1\textwidth,clip,angle=0]{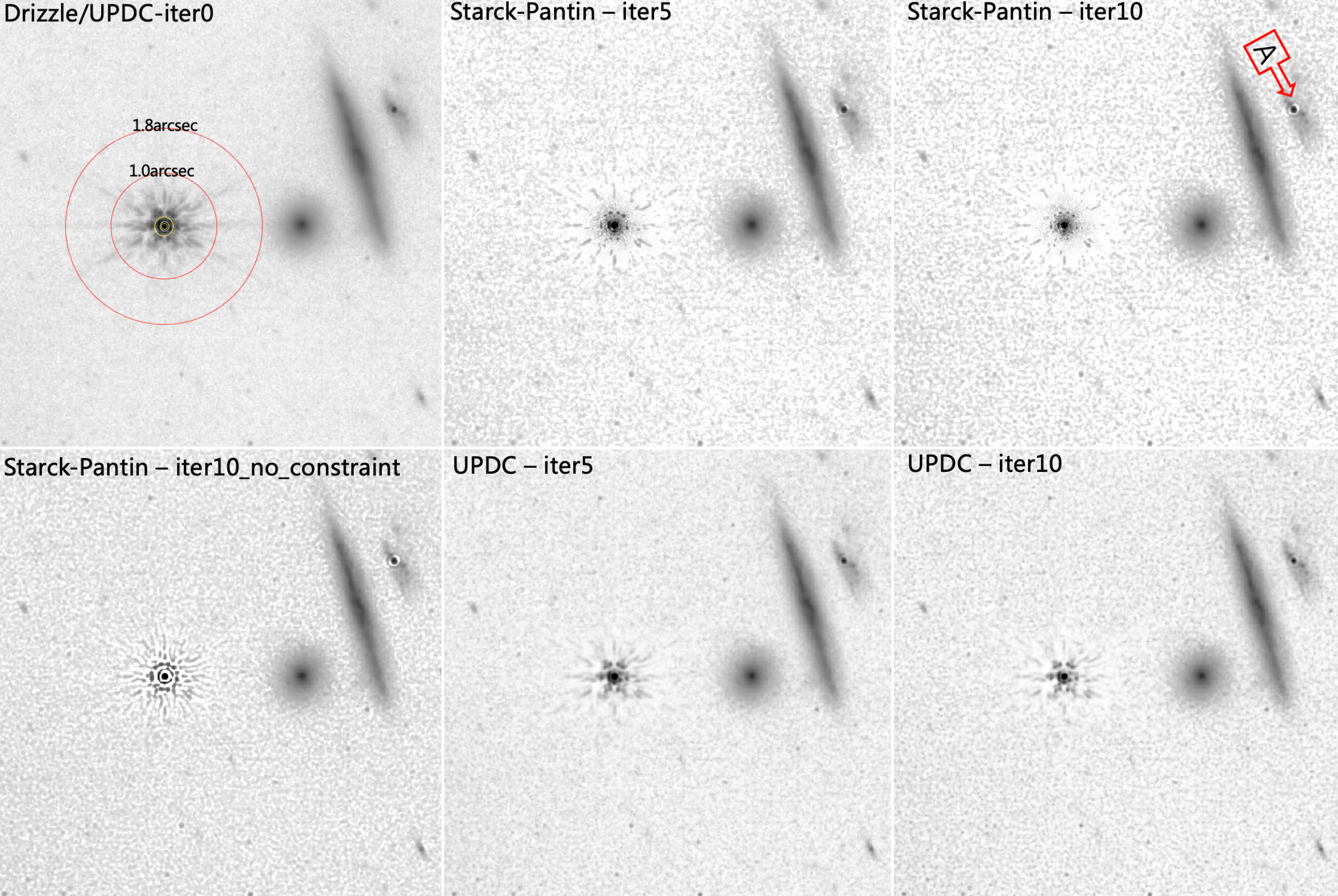}
\caption{{Coaddition for a point source from SMACSJ0723 (a star at $R.A.=110^\circ.865243, DEC=-73^\circ.454731$) using different methods in the F200W band. Five circles are displayed on the top left panel, representing radii of 50mas, 80mas (radius of first minimum), 150mas, 1arcsec, and 1.8arcsec, respectively. We primarily compare our pipeline with the Stark-Pantin algorithm at the 5th and 10th iterations under the same positivity constraint (equation \ref{cstr}). If the constraint is removed, Stark-Pantin coaddition produces a significant ringing effect around bright point sources as shown on the bottom left panel. {\bf Tag-A} denotes another point source which is severely affected by the ringing effect during the Stark-Pantin coaddition. The UPDC algorithm controls noise well and reduces the ringing effect significantly under the same number of iteration.}}
\label{compfig}
\end{figure*}

\item{We have investigated to determine the dependency of faint source detection upon UPDC iterations. Figure \ref{det_rate} shows that the number of detections increases as iterations increase. In this study, all stars are masked from the parallel field of SMACSJ0723. The left panel of the figure displays the total detection numbers for the three LW bands, whereas the right panel shows the corresponding detection growth rates, which are normalized to the detection number of the Drizzle image (UPDC-iter0). Based on the detection growth curve in Figure \ref{det_rate}, it is evident that the increase in the number of new detections from the 5th to the 10th iteration is not as significant as from the 0th to the 5th iteration. As illustrated in the bottom row of panels of Figure \ref{detection}, compared to UPDC-iter5 coadditions, there are only 4 new detections in the UPDC-iter10 images. }

\item{The UPDC algorithm has a significant effect not only on the detection of faint sources but also on the aperture photometry of those sources. To better understand the effect of UPDC on aperture photometry, we selected a sample of sources that were cross-matched in both the UPDC-iter0 and UPDC-iter5 coaddition\footnote{To balance the signal extraction and noise amplification, we believe that the UPDC-iter5 is more suitable for detecting the underlying faint sources.} reconstructed images of UPDC. We used the $\rm SourceExtractor$'s automatic aperture magnitude ($\rm MAG_{AUTO}$) routine\citep{1996A&AS..117..393B} to measure the photometry of these sources. This routine is based on Kron's "first-moment" algorithm \citep{Kron_1980} and is known to provide accurate estimates of "total magnitudes", particularly for galaxies.

Figure \ref{fig_mag_comp} shows the photometry of the cross-matched sources (UPDC-iter0 vs. UPDC-iter5) from the parallel field of SMACSJ0723. As a result, the magnitudes of sources in UPDC-iter5 are higher than those in UPDC-iter0 (Drizzle). This difference is evident for the vast majority of the sources, especially at the faint end. Around magnitude 28, the mean of UPDC-iter0 and UPDC-iter5 diverges by about $0.3-0.5$ magnitude. For the LW channel 4.0 $\mu$m, an increase of 0.4 magnitudes indicates that the UPDC has recovered the flux (30\% of the total) that had spread beyond the first minimum of the "snowflake" pattern (radius $\sim0\arcsec.16$) due to the PSF blurring (see the JWST official document "JWST-STScI-001157.pdf"\footnote{\url{https://www.stsci.edu/files/live/sites/www/files/home/jwst/documentation/technical-documents/_documents/JWST-STScI-001157.pdf}}). The majority of the points lie above the diagonal line, which is mainly due to the deblending effect brought about by UPDC, as indicated by the green and orange arrows in Figure \ref{detection}.

If there are multiple sources located below the detection threshold (6$\sigma_{\rm bkg}$) of $\rm SourceExtractor$ in the vicinity of a faint source (UPDC-iter0), then after UPDC processing (UPDC-iter5), these sources may become detectable and blend with the existing faint source, leading to a significant increase in the magnitude. These are the cases where the deviation is significantly below the diagonal line. At the bright end, both the UPDC and Drizzle algorithms exhibit similar performance and tend to converge.}

\begin{figure*}
\centering
\includegraphics[width=1\textwidth,clip,angle=0]{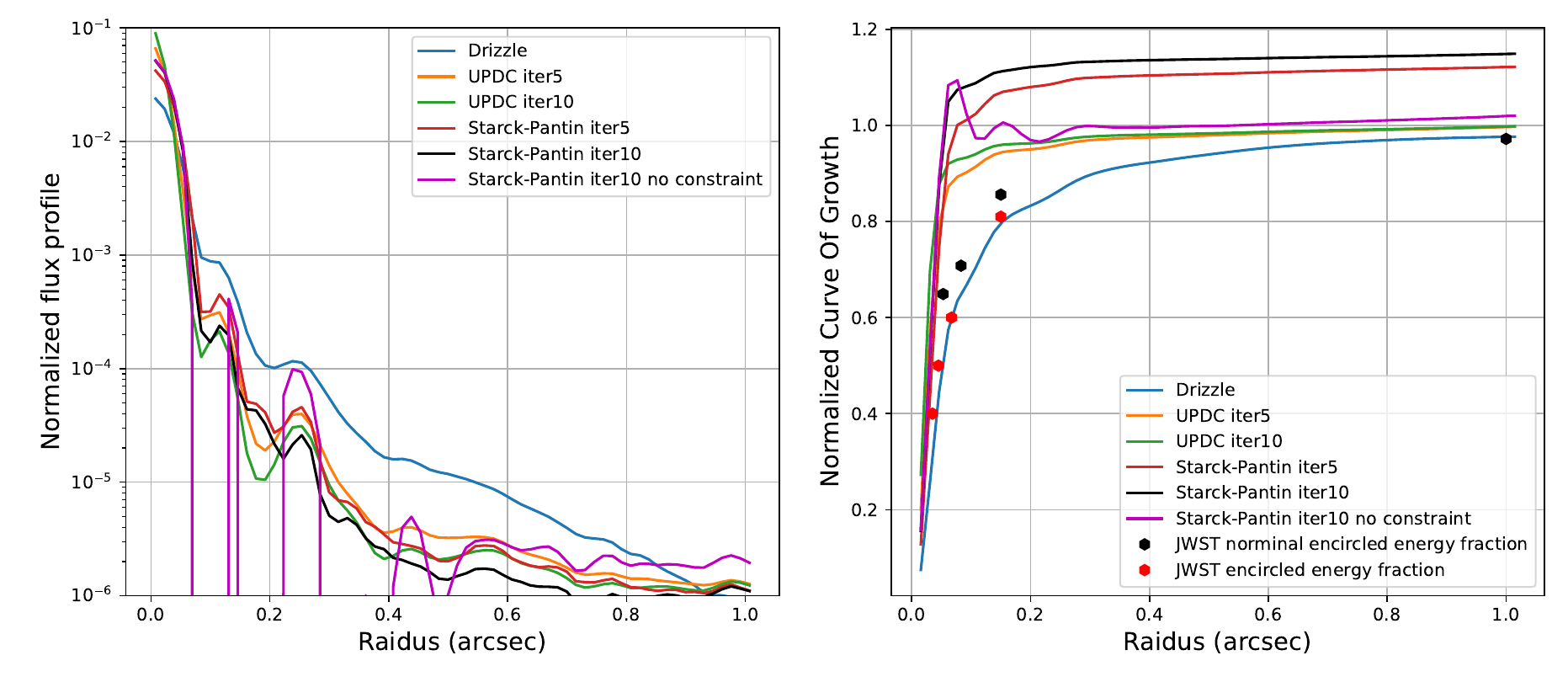}
\caption{Normalized flux profile (left panel) and curve of growth (right panel) for a star at $R.A.=110^\circ.865243, DEC=-73^\circ.454731$. The measurements are obtained from the coaddition of SMACSJ0723 exposures in the F200W band. On the left panel, all curves represent the flux normalized to the total flux within circle of radius 1 arcsec ($\simeq$ 67 pixels). There are several valleys in each curve, which corresponds to the minimums of the Airy pattern. Black (red) hexagons on the right panel indicate the nominal (the officially measured) encircled energy fraction at different radii. All curves on the right panel are normalized to the total flux measured on the Drizzled image within circle of radius 120 pixels ($\simeq 1.8$ arcsec).}
\label{profile}
\end{figure*}

\begin{figure*}
\centering
\includegraphics[width=1\textwidth,clip,angle=0]{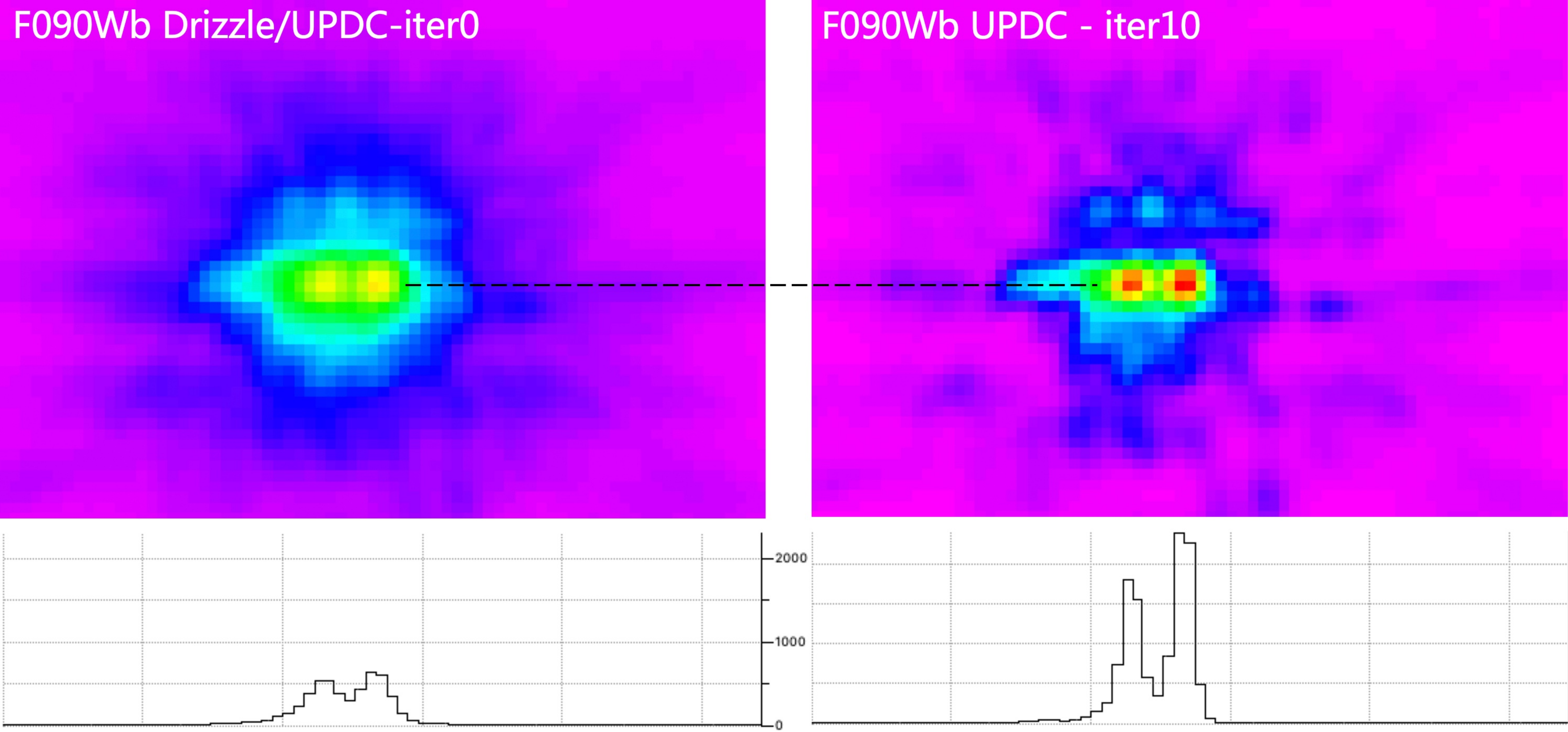}
\caption{A closed binary is resolved by the UPDC in the SMACSJ0723 field. The binary is located at $R.A.=110^\circ.845109, DEC=-73^\circ.455443$. The lower panels present a cross-sectional view of the upper images along the dashed line. The binary in the left lower image is not fully resolved, but the one in the right lower image provides significant evidence.}
\label{binary_resolved}
\end{figure*}

\item{{The UPDC method provides better fidelity, sharper contours, and more accurate aperture photometry when reconstructing point sources compared to other methods like the Stark-Pantin method. Figure \ref{compfig} displays a series of coadditions for a point source from SMACSJ0723 (a star at $R.A.=110^\circ.865243, DEC=-73^\circ.454731$) using different methods in the F200W band. Five circles on the top left panel represent radii of 50mas, 80mas (radius of first minimum), 150mas, 1arcsec, and 1.8arcsec, respectively. Our primary comparison is between the UPDC algorithm and the Stark-Pantin algorithm at the 5th and 10th iterations under the same positivity constraint (equation \ref{cstr}). If the constraint is removed, the Stark-Pantin coaddition produces a significant ringing effect around bright point sources, as shown in the bottom left panel. {\bf Tag-A} represents another point source significantly affected by the ringing effect introduced by the Stark-Pantin method. The UPDC algorithm effectively manages noise and reduces the ringing effect under the condition of the same number of iterations.}

Quantitative comparisons are illustrated in Figure \ref{profile}. Using "Profiles" ($photutils.profiles$) in the software $Photutils$, we have calculated the normalized flux profile (left panel) and the curve of growth (encircled energy, right panel) for the star with coordinates $R.A.=110^\circ.865243, DEC=-73^\circ.454731$. The left panel displays curves that represent the flux normalized to the total flux within a circle of radius 1 arcsec (approximately 67 pixels). It is evident that after the coaddition, our pipeline produces a sharper profile compared to the Starck-Pantin method. Beyond a radius of 0.7 arcsec, the profiles obtained from Starck-Pantin and UPDC methods tend to flatten out, indicating that beyond this radius, the signal from the point source has been overwhelmed by noise. While the profile from Drizzle shows a significant drop, indicating that there is still a strong signal that has not been extracted, also see the top left panel of Figure \ref{compfig}. Black (red) hexagons on the right panel indicate the nominal (the officially measured) encircled energy fraction at different radii (see table 1, 2 and page 27 of the JWST official document "JWST-STScI-001157.pdf"). All curves on the right panel are normalized to the total flux measured on the Drizzled image within a circle of radius 120 pixels (approximately 1.8 arcsecs, flux percentage $\sim99.9\%$ for the WebbPSF/HybPSF). For the Starck-Pantin method, the non-strict positivity constraint (equation \ref{cstr}) can significantly reduce the ringing effect around the bright star. However, the constraint may not effectively preserve flux. 
}

\item{
{According to the Fraunhofer dispersion of light, the radii of the minimums of the Airy disk are proportional to the wavelength of the incident light. The LW light in polychromatic light will be diffracted to the periphery, while the SW light will remain at the centre, which is evident in the RGB Figures \ref{SM0723-1} to \ref{PF-1}. This will make the source appear bluer. The light dispersed by the diffraction effect is collected back to form polychromatic light, which is the inverse dispersion effect. Like other PSF deconvolution methods, UPDC can achieve inverse dispersion, which has better effects. As shown in Figure \ref{compfig}, compared to the stacking results of Starck-Pantin, UPDC maintains a self-similar structure during the iteration process (indicating that it retains diffraction information), which is the expected case of a good iteration method. From the left panel of Figure \ref{profile}, it is obvious that after the same number of iterations, the point source reconstructed by UPDC has a higher peak flux, and a higher concentration of the surface brightness, indicating its PSF deconvolution effect (including the inverse dispersion) is better. As shown in Figures \ref{SM0723-1} to \ref{PF-1}, many blue point-like sources that were originally present in the Drizzle images have now turned into white ones.  This inverse dispersion effect provides more accurate colours to these sources, which could prove to be useful while searching for faint sources in multi-wavelength stacked images.}
}

\item{ Upsampling and PSF deconvolution can improve the spatial resolution of the target image beyond the optical diffraction limit of the JWST. We detected a binary star in the vicinity of SMACSJ0723 in the F090W band via our pipeline. As shown in the left panels of Figure \ref{binary_resolved}, the binary is not completely resolved yet\footnote{Note that it is unresolved in the exposure image.}. However, in the right panels, the binary stars can be clearly distinguished. The lower panels illustrate a cross-sectional view along the dashed line of the upper images. The binary can be found at $R.A.=110^\circ.845109, DEC=-73^\circ.455443$. The HST's optical band observations have provided additional evidence of this binary star system. GAIA's observations suggest that the system consists of an $M$-type star located at a distance of around 1 kpc from us. The system has a mass ratio of roughly $1:1$ and a separation of approximately 60 AU, making it a close binary system.}

\end{enumerate}

{By inferring the image that gets convolved with the PSF and pixel sampling to approach the real exposures, we have developed the UPDC, which is a scientific data pipeline module initially designed for the CSST-MCI. By using UPDC, we have achieved exceptional results in the coaddition of the multiple exposures from JWST.}

\section{SUMMARY}
\label{sec:summ}
 Our pipeline not only delivers image enhancement with higher fidelity but also exhibits significant improvements in the detection of faint sources, photometry, and the deblending (super-resolution) of closely packed sources compared to the previous algorithms (Drizzle / Starck-Pantin). Based on the tests mentioned in section \ref{sec:results}, we have drawn the following conclusions.
\begin{enumerate}[label=(\roman*).]

\item{
Multiple exposures coadded by our pipeline are visually clearer and sharper, enabling us to detect more details such as the substructures of spiral galaxies, objects hidden by neighboring bright sources, various components (dust) of galaxies, and closed binaries, as shown on Figure \ref{SM0723-1} - \ref{PF-1} and Figure \ref{binary_resolved}. This is mainly due to the under-sampled exposures anti-aliasing and super-resolution capabilities provided by the UPDC algorithm.
}

\item{
By suppressing the PSF, the flux of the source becomes more concentrated on the source itself. This results in better photometric accuracy, especially in crowded fields like those found around galaxy clusters and nearby galaxies and globular clusters. The UPDC algorithm also makes it easier for people to deblend overlapped sources on images, and higher accuracy can be obtained by aperture photometry for such sources.
}

\item{
The UPDC software incorporates a PSF deconvolution algorithm that may assist in enhancing the signal-to-noise ratio of faint sources, like high-redshift galaxies or quasars. This algorithm enables the original flux of the source, which was initially spread by PSF, to be recollected, thereby facilitating the detection of such sources. By improving the SNR, this algorithm can enhance the accuracy of measurements and enable researchers to acquire more precise data.
}

\item{The UPDC reconstruction, such as UPDC-iter5 and its following iterations, improve the optical resolution by at least a factor of two in all six bands compared to the Drizzle coadded images of JWST. This is crucial for observing detailed structures of distant objects, such as resolving quasars from their host galaxies and distinguishing the substructures of high-redshift galaxies. The under-sampled exposures are anti-aliased completely, particularly in the F090W, F150W, F277W and F356W bands.}

\end{enumerate}

In this work, the JWST images processed by the UPDC are called the "UPdec-Webb" dataset and released on the official website of China-VO: \dataset[doi:10.12149/101436]{https://nadc.china-vo.org/res/r101436/}. To fulfil the varied scientific goals of researchers, each coaddition consists of 10 iterations. We will first release the UPDC stacked images of the SMACSJ0723 field (JWST-module B), and then its parallel field (JWST-module A). After further upgrading and refining the program, especially in its ability to effectively estimate uncertainties and weights, we may release the source code of UPDC in our subsequent articles.

\section{Acknowledgments}
{We are grateful for the valuable comments provided by the anonymous reviewer, which have greatly improved the presentation of this paper. In particular, his/her professionalism, patience, attention to detail, and insight into the issues have profoundly enhanced the potential impact of this work, and we express our special thanks to the reviewer.} This work is supported by the Foundation for Distinguished Young Scholars of Jiangsu Province (No. BK20140050), the National Natural Science Foundation of China (Nos. 11973070, 11333008, 11273061, 11825303, and 11673065), the China Manned Space Project with No. CMS-CSST-2021-A01, CMS-CSST-2021-A03, CMS-CSST-2021-B01, the GHfund A (202302017475) and the Joint Funds of the National Natural Science Foundation of China (No. U1931210). HYS acknowledges the support from the Key Research Program of Frontier Sciences, CAS, Grant No. ZDBS-LY-7013 and Program of Shanghai Academic/Technology Research Leader. We acknowledge the support from the science research grants from the Ministry of Science and Technology of China (grant Nos. 2020SKA0110100), the China Manned Space Project with CMS-CSST-2021-A04, and CMS-CSST-2021-A07.

The Early Release Observations and associated materials were developed, executed, and compiled by the ERO production team:  Hannah Braun, Claire Blome, Matthew Brown, Margaret Carruthers, Dan Coe, Joseph DePasquale, Nestor Espinoza, Macarena Garcia Marin, Karl Gordon, Alaina Henry, Leah Hustak, Andi James, Ann Jenkins, Anton Koekemoer, Stephanie LaMassa, David Law, Alexandra Lockwood, Amaya Moro-Martin, Susan Mullally, Alyssa Pagan, Dani Player, Klaus Pontoppidan, Charles Proffitt, Christine Pulliam, Leah Ramsay, Swara Ravindranath, Neill Reid, Massimo Robberto, Elena Sabbi, Leonardo Ubeda. The EROs were also made possible by the foundational efforts and support from the JWST instruments, STScI planning and scheduling, and Data Management teams.

{The JWST data presented in this article were obtained from the MAST at the Space Telescope Science Institute. The JWST officially released Stage-3 coaddition ($*i2d.fits$) shown in the left panel of Figure \ref{mastdb-updc} can be accessed via \dataset[doi: 10.17909/7wqf-xd87]{https://doi.org/10.17909/7wqf-xd87}. The Stage-2 exposures ($*cal.fits$) can be accessed via MAST search\footnote{\url{https://mast.stsci.edu/search/ui/\#/jwst/results?resolve=true&instruments=NIRCAM&program_id=2736&useStore=false&search_key=7055aba0ea26b8} }.

 In this study, a cluster is used with the SIMT accelerator made in China. The cluster includes many nodes each containing 2 CPUs and 8 accelerators. The accelerator adopts a GPU-like architecture consisting of a 64GB HBM2 device memory and many compute units. Accelerators connected to CPUs with PCI-E, the peak bandwidth of the data transcription between main memory and device memory is 64GB/s.

\facilities{JWST(NIRCam), HST(ACS), GAIA.}

\software{{astropy \citep{2013A&A...558A..33A,2018AJ....156..123A},
          SourceExtractor \citep{1996A&AS..117..393B,2010ascl.soft10064B},
          JWST-pipeline \citep{2017ASPC..512..355B,2019ASPC..523..543B,2022zndo...7038885B},
          HybPSF \citep{2024AJ....167...58N}, 
          Photutils \citep{larry_bradley_2024_12585239},
          Drizzlepac \citep{2010bdrz.conf..382F,2021AAS...23821602H},
          UPDC \citep{Wang_2022}.}
          }

\appendix{}
\section{Uncertainty array} \label{errors}
{The uncertainty calculation is a complex task in the forward modeling algorithm, involving sub-pixel operation and PSF convolution. However, the JWST-pipeline document offers an approximate method for evaluating uncertainties on the Drizzled image\footnote{\url{https://jwst-pipeline.readthedocs.io/en/stable/jwst/resample/main.html}}. This means that error propagation of Drizzle can be implemented using the Drizzle method itself. Following this approach, if we perform dithering, PSF convolution, and resampling (matching those used for the science data) on an underlying uncertainty array to obtain noise images comparable to those in real exposures (the extensions in the $*cal.fits$ files), then we can consider the output uncertainty array as a reasonable error estimation for the UPDC images. Our steps are shown as follows.}

\begin{enumerate}[label=\textbf{Step \arabic*.}]

\item {For each Stage-2 image ($*cal.fits$), take the square root of the read noise variance array (the VAR\_RNOISE extension) to make an error image. Instead of the real multiple exposures, here we use the error images as the UPDC's input images.}

\item {For the UPDC-iter0 image, also take the square root of the read noise variance array (the VAR\_RNOISE extension in the Stage-3 image $*i2d.fits$) to create an error array as the initial condition for the UPDC error estimation.}

\item {Run UPDC to coadd the error images, with UPDC parameters and HybPSFs matching those used for the science data. It is important to note two differences from the JWST-pipeline method in the UPDC processing. First, when we reach the PSF convolution step, we square the error to turn it into a variance, then convolve the variance with the PSF. After the PSF convolution, we take the square root of the variance to convert it back into an error for the subsequent resampling steps. Our numerical simulations show that, at least for Poisson noise, the uncertainty estimated in this way is comparable to the truth, as shown in Figure \ref{figerrors}. Second, when calculating the ratio of real observation errors to mock observation errors, i.e. $\frac{g_k}{ \Dd^h_k\{ \Bb_k\{f^{(i)}\} \}}$ in Equation \ref{WLMPHRC}, we perform a classical error propagation rather than simply taking the ratio of the two. }

\item {Square the outputs and obtain the read noise variance array for each iteration.}

\item {Perform the same steps for the Poisson noise variance and the flat variance. }

\end{enumerate}

{Due to different uncertainty sources behave differently, after UPDC processes each variance component, the final error array 'ERR' is computed as the square root of the sum of the three independent variance components. }

\begin{figure*}
\centering
\includegraphics[width=0.8\textwidth,clip]{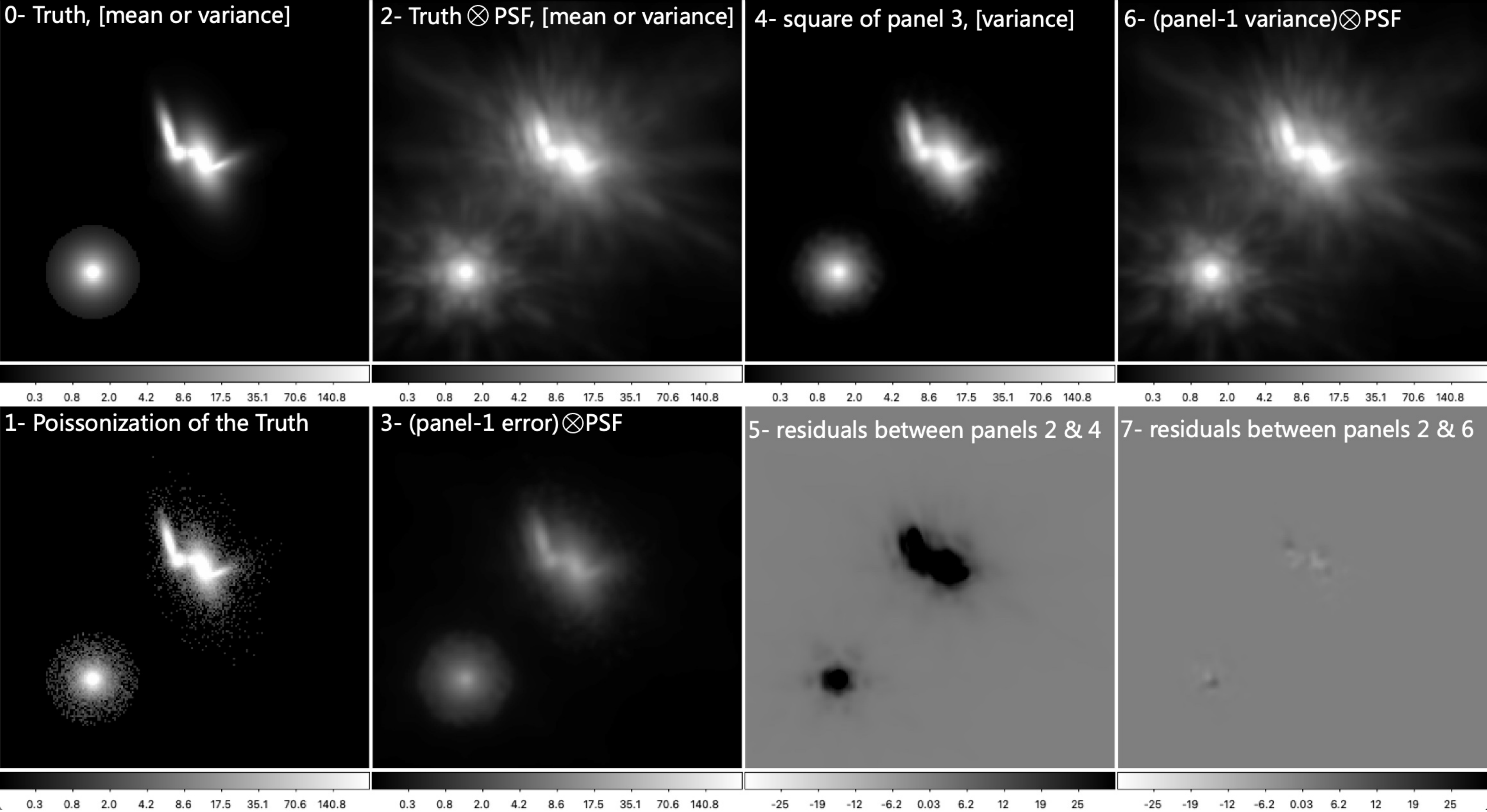}
\caption{Error propagation of PSF convolution in simulation with Poisson noise. Panel 0 shows the underlying image (truth). For a Poisson distribution, the mean and variance are equal. To generate Panel 1, we Poissonize the truth (Panel 0). Panel 2 represents the result of convolving the truth with a PSF (the ideal case). By taking the square root of Panel 1 and convolving it with the same PSF, we obtain Panel 3 (the convolved error). Squaring Panel 3 gives us Panel 4 (the variance). Convolution of the variance from Panel 1 with the same PSF results in Panel 6. Meanwhile, Panels 5 and 7 represent the residuals for these two types of error propagation. This simulation indicates that calculating error propagation of PSF convolution through variance is more accurate than doing so through the error itself.}
\label{figerrors}
\end{figure*}

{The iterations of UPDC can simultaneously enhance the signal and amplify the noise, with the SNR rising quickly at first and then slowly decreasing\citep{Starck+2002, Wang_2022}. In the previous numerical simulation tests, when coadding multiple exposures with predominantly Gaussian/Poisson noise, UPDC exhibits superior control over the noise on the whole image compared to other methods\citep{Wang_2022} such as Starck-Pantin, Landweber, and Richardson-Lucy. If no noise is introduced in the initial exposures, the iterations provided by UPDC (or Starck-Pantin) are convergent and converge to the underlying image (the ground truth). This indicates that throughout the iterative process, these methods consistently aim to identify all signals in the residual image. However, due to the degeneracy between the noise introduced by exposure and the actual signals, the algorithm mistakenly amplifies noise fluctuations as if they were signals. In this sense, image coadding algorithms do not change the nature of the noise but rather alter its amplitude. This insight will aid us in quickly estimating the noise level on each iteration output, a topic that we will further explore and validate by the numerical simulations in subsequent research work.
}

\section{weight}\label{weights}
Generally, weights for each iteration output and each input exposure can be evaluated by its inverse variance map. For the UPDC-iter0, the Drizzle has already calculated weight and flux \citep{Fruchter+2002}. From UPDC-iter$N$ to mock exposures, the $Blot$ procedure has already considered the weights. However, for the PSF convolution and the drizzling of comparison residual, we have to assume all pixels (input/mock/output) are weighted equally. Since $(i)$ it is quite challenging to take the weight into account in PSF convolution; $(ii)$ we have effectively addressed issues such as cosmic rays and jumping pixels by replacing them with interpolations or smoothing the jumps; $(iii)$ for a ratio correction term in the iteration, e.g. Eq. \ref{WLMPHRC}, the weight terms in the numerator and denominator cancel each other out [see also chapter 5.4 in the previous work \citep{Starck+2002}], we roughly treat all pixels as equally weighted to simplify the computation for PSF convolution and residual drizzling steps.

On the other hand, due to the impact of read noise and flat field noise on UPDC outputs not being well evaluated in our numerical simulations, we have not attached the weight map to the current version of the released dataset. This is just a temporary solution. We will add a weight "WHT" extension (also "ERR", "CON", "VAR\_POISSON", "VAR\_RNOISE", "VAR\_FLAT", "HDRTAB" and "ASDF\_METADATA" extensions) to the $fits$ files in subsequent data upgrades and abolish the interpolation of abnormal pixels, retaining them as “NaN” instead.

\clearpage\newpage
\bibliography{JWSTcoadd}{}

\begin{thebibliography}{}
\expandafter\ifx\csname natexlab\endcsname\relax\def\natexlab#1{#1}\fi
\providecommand{\url}[1]{\href{#1}{#1}}
\providecommand{\dodoi}[1]{doi:~\href{http://doi.org/#1}{\nolinkurl{#1}}}
\providecommand{\doeprint}[1]{\href{http://ascl.net/#1}{\nolinkurl{http://ascl.net/#1}}}
\providecommand{\doarXiv}[1]{\href{https://arxiv.org/abs/#1}{\nolinkurl{https://arxiv.org/abs/#1}}}

\bibitem[{{Astropy Collaboration} {et~al.}(2013){Astropy Collaboration},
  {Robitaille}, {Tollerud}, {Greenfield}, {Droettboom}, {Bray}, {Aldcroft},
  {Davis}, {Ginsburg}, {Price-Whelan}, {Kerzendorf}, {Conley}, {Crighton},
  {Barbary}, {Muna}, {Ferguson}, {Grollier}, {Parikh}, {Nair}, {Unther},
  {Deil}, {Woillez}, {Conseil}, {Kramer}, {Turner}, {Singer}, {Fox}, {Weaver},
  {Zabalza}, {Edwards}, {Azalee Bostroem}, {Burke}, {Casey}, {Crawford},
  {Dencheva}, {Ely}, {Jenness}, {Labrie}, {Lim}, {Pierfederici}, {Pontzen},
  {Ptak}, {Refsdal}, {Servillat}, \& {Streicher}}]{2013A&A...558A..33A}
{Astropy Collaboration}, {Robitaille}, T.~P., {Tollerud}, E.~J., {et~al.} 2013,
  \aap, 558, A33, \dodoi{10.1051/0004-6361/201322068}

\bibitem[{{Astropy Collaboration} {et~al.}(2018){Astropy Collaboration},
  {Price-Whelan}, {Sip{\H{o}}cz}, {G{\"u}nther}, {Lim}, {Crawford}, {Conseil},
  {Shupe}, {Craig}, {Dencheva}, {Ginsburg}, {VanderPlas}, {Bradley},
  {P{\'e}rez-Su{\'a}rez}, {de Val-Borro}, {Aldcroft}, {Cruz}, {Robitaille},
  {Tollerud}, {Ardelean}, {Babej}, {Bach}, {Bachetti}, {Bakanov}, {Bamford},
  {Barentsen}, {Barmby}, {Baumbach}, {Berry}, {Biscani}, {Boquien}, {Bostroem},
  {Bouma}, {Brammer}, {Bray}, {Breytenbach}, {Buddelmeijer}, {Burke},
  {Calderone}, {Cano Rodr{\'\i}guez}, {Cara}, {Cardoso}, {Cheedella}, {Copin},
  {Corrales}, {Crichton}, {D'Avella}, {Deil}, {Depagne}, {Dietrich}, {Donath},
  {Droettboom}, {Earl}, {Erben}, {Fabbro}, {Ferreira}, {Finethy}, {Fox},
  {Garrison}, {Gibbons}, {Goldstein}, {Gommers}, {Greco}, {Greenfield},
  {Groener}, {Grollier}, {Hagen}, {Hirst}, {Homeier}, {Horton}, {Hosseinzadeh},
  {Hu}, {Hunkeler}, {Ivezi{\'c}}, {Jain}, {Jenness}, {Kanarek}, {Kendrew},
  {Kern}, {Kerzendorf}, {Khvalko}, {King}, {Kirkby}, {Kulkarni}, {Kumar},
  {Lee}, {Lenz}, {Littlefair}, {Ma}, {Macleod}, {Mastropietro}, {McCully},
  {Montagnac}, {Morris}, {Mueller}, {Mumford}, {Muna}, {Murphy}, {Nelson},
  {Nguyen}, {Ninan}, {N{\"o}the}, {Ogaz}, {Oh}, {Parejko}, {Parley}, {Pascual},
  {Patil}, {Patil}, {Plunkett}, {Prochaska}, {Rastogi}, {Reddy Janga},
  {Sabater}, {Sakurikar}, {Seifert}, {Sherbert}, {Sherwood-Taylor}, {Shih},
  {Sick}, {Silbiger}, {Singanamalla}, {Singer}, {Sladen}, {Sooley},
  {Sornarajah}, {Streicher}, {Teuben}, {Thomas}, {Tremblay}, {Turner},
  {Terr{\'o}n}, {van Kerkwijk}, {de la Vega}, {Watkins}, {Weaver}, {Whitmore},
  {Woillez}, {Zabalza}, \& {Astropy Contributors}}]{2018AJ....156..123A}
{Astropy Collaboration}, {Price-Whelan}, A.~M., {Sip{\H{o}}cz}, B.~M., {et~al.}
  2018, \aj, 156, 123, \dodoi{10.3847/1538-3881/aabc4f}

\bibitem[{Atek {et~al.}(2022)Atek, Shuntov, Furtak, Richard, Kneib, Mahler,
  Zitrin, McCracken, Charlot, Chevallard, \&
  Chemerynska}]{10.1093/mnras/stac3144}
Atek, H., Shuntov, M., Furtak, L.~J., {et~al.} 2022, Monthly Notices of the
  Royal Astronomical Society, 519, 1201, \dodoi{10.1093/mnras/stac3144}

\bibitem[{{Bagley} {et~al.}(2023){Bagley}, {Finkelstein}, {Koekemoer},
  {Ferguson}, {Arrabal Haro}, {Dickinson}, {Kartaltepe}, {Papovich},
  {P{\'e}rez-Gonz{\'a}lez}, {Pirzkal}, {Somerville}, {Willmer}, {Yang}, {Yung},
  {Fontana}, {Grazian}, {Grogin}, {Hirschmann}, {Kewley}, {Kirkpatrick},
  {Kocevski}, {Lotz}, {Medrano}, {Morales}, {Pentericci}, {Ravindranath},
  {Trump}, {Wilkins}, {Calabr{\`o}}, {Cooper}, {Costantin}, {de la Vega},
  {Hilbert}, {Hutchison}, {Larson}, {Lucas}, {McGrath}, {Ryan}, {Wang}, \&
  {Wuyts}}]{Bagley2023ApJ...946L..12B}
{Bagley}, M.~B., {Finkelstein}, S.~L., {Koekemoer}, A.~M., {et~al.} 2023,
  \apjl, 946, L12, \dodoi{10.3847/2041-8213/acbb08}

\bibitem[{Bates \& Cady(1980)}]{Bates+1980}
Bates, R., \& Cady, F. 1980, Optics Communications, 32, 365,
  \dodoi{https://doi.org/10.1016/0030-4018(80)90261-8}

\bibitem[{{Bertin} \& {Arnouts}(1996)}]{1996A&AS..117..393B}
{Bertin}, E., \& {Arnouts}, S. 1996, \aaps, 117, 393,
  \dodoi{10.1051/aas:1996164}

\bibitem[{{Bertin} \& {Arnouts}(2010)}]{2010ascl.soft10064B}
---. 2010, {SExtractor: Source Extractor}, Astrophysics Source Code Library,
  record ascl:1010.064.
\newblock \doeprint{1010.064}

\bibitem[{Bradley {et~al.}(2024)Bradley, Sip{\H o}cz, Robitaille, Tollerud,
  Vin{\'{\i}}cius, Deil, Barbary, Wilson, Busko, Donath, G{\"u}nther, Cara,
  Lim, Me{\ss}linger, Burnett, Conseil, Droettboom, Bostroem, Bray, Bratholm,
  Jamieson, Ginsburg, Barentsen, Craig, Pascual, Rathi, Perrin, Morris, \&
  Perren}]{larry_bradley_2024_12585239}
Bradley, L., Sip{\H o}cz, B., Robitaille, T., {et~al.} 2024, astropy/photutils:
  1.13.0, 1.13.0,  Zenodo, \dodoi{10.5281/zenodo.12585239}

\bibitem[{{Bushouse} {et~al.}(2017){Bushouse}, {Droettboom}, \&
  {Greenfield}}]{2017ASPC..512..355B}
{Bushouse}, H., {Droettboom}, M., \& {Greenfield}, P. 2017, in Astronomical
  Society of the Pacific Conference Series, Vol. 512, Astronomical Data
  Analysis Software and Systems XXV, ed. N.~P.~F. {Lorente}, K.~{Shortridge},
  \& R.~{Wayth}, 355

\bibitem[{{Bushouse} {et~al.}(2019){Bushouse}, {Eisenhamer}, \&
  {Davies}}]{2019ASPC..523..543B}
{Bushouse}, H., {Eisenhamer}, J., \& {Davies}, J. 2019, in Astronomical Society
  of the Pacific Conference Series, Vol. 523, Astronomical Data Analysis
  Software and Systems XXVII, ed. P.~J. {Teuben}, M.~W. {Pound}, B.~A.
  {Thomas}, \& E.~M. {Warner}, 543

\bibitem[{{Bushouse} {et~al.}(2022){Bushouse}, {Eisenhamer}, {Dencheva},
  {Davies}, {Greenfield}, {Morrison}, {Hodge}, {Simon}, {Grumm}, {Droettboom},
  {Slavich}, {Sosey}, {Pauly}, {Miller}, {Jedrzejewski}, {Hack}, {Davis},
  {Crawford}, {Law}, {Gordon}, {Regan}, {Cara}, {MacDonald}, {Bradley},
  {Shanahan}, \& {Jamieson}}]{2022zndo...7038885B}
{Bushouse}, H., {Eisenhamer}, J., {Dencheva}, N., {et~al.} 2022, {JWST
  Calibration Pipeline}, 1.7.0, Zenodo,  Zenodo, \dodoi{10.5281/zenodo.7038885}

\bibitem[{Coe {et~al.}(2019)Coe, Salmon, Bradač, Bradley, Sharon, Zitrin,
  Acebron, Cerny, Cibirka, Strait, Paterno-Mahler, Mahler, Avila, Ogaz, Huang,
  Pelliccia, Stark, Mainali, Oesch, Trenti, Carrasco, Dawson, Rodney, Strolger,
  Riess, Jones, Frye, Czakon, Umetsu, Vulcani, Graur, Jha, Graham, Molino,
  Nonino, Hjorth, Selsing, Christensen, Kikuchihara, Ouchi, Oguri, Welch,
  Lemaux, Andrade-Santos, Hoag, Johnson, Peterson, Past, Fox, Agulli,
  Livermore, Ryan, Lam, Sendra-Server, Toft, Lovisari, \& Su}]{Coe_2019}
Coe, D., Salmon, B., Bradač, M., {et~al.} 2019, The Astrophysical Journal,
  884, 85, \dodoi{10.3847/1538-4357/ab412b}

\bibitem[{{Ding} {et~al.}(2022){Ding}, {Silverman}, \&
  {Onoue}}]{2022ApJ...939L..28D}
{Ding}, X., {Silverman}, J.~D., \& {Onoue}, M. 2022, \apjl, 939, L28,
  \dodoi{10.3847/2041-8213/ac9c02}

\bibitem[{Farsiu {et~al.}(2004)Farsiu, Robinson, Elad, \&
  Milanfar}]{Farsiu+2004a}
Farsiu, S., Robinson, M., Elad, M., \& Milanfar, P. 2004, IEEE Transactions on
  Image Processing, 13, 1327, \dodoi{10.1109/TIP.2004.834669}

\bibitem[{Fruchter(2011)}]{Fruchter+2011}
Fruchter, A.~S. 2011, Publications of the Astronomical Society of the Pacific,
  123, 497, \dodoi{10.1086/659313}

\bibitem[{{Fruchter} \& {et al.}(2010)}]{2010bdrz.conf..382F}
{Fruchter}, A.~S., \& {et al.} 2010, in 2010 Space Telescope Science Institute
  Calibration Workshop, 382--387

\bibitem[{Fruchter \& Hook(2002)}]{Fruchter+2002}
Fruchter, A.~S., \& Hook, R.~N. 2002, Publications of the Astronomical Society
  of the Pacific, 114, 144, \dodoi{10.1086/338393}

\bibitem[{{Hoffmann} {et~al.}(2021){Hoffmann}, {Mack}, {Avila}, {Martlin},
  {Cohen}, \& {Bajaj}}]{2021AAS...23821602H}
{Hoffmann}, S.~L., {Mack}, J., {Avila}, R., {et~al.} 2021, in American
  Astronomical Society Meeting Abstracts, Vol.~53, American Astronomical
  Society Meeting Abstracts, 216.02

\bibitem[{Irani \& Peleg(1993)}]{Irani+1993}
Irani, M., \& Peleg, S. 1993, J Vis. Commun. Image Represent, 4, 324

\bibitem[{{Kron}(1980)}]{Kron_1980}
{Kron}, R.~G. 1980, \apjs, 43, 305, \dodoi{10.1086/190669}

\bibitem[{{Lucy}(1974)}]{1974AJ.....79..745L}
{Lucy}, L.~B. 1974, \aj, 79, 745, \dodoi{10.1086/111605}

\bibitem[{Montes \& Trujillo(2022)}]{Montes_2022}
Montes, M., \& Trujillo, I. 2022, The Astrophysical Journal Letters, 940, L51,
  \dodoi{10.3847/2041-8213/ac98c5}

\bibitem[{Morishita \& Stiavelli(2023)}]{Morishita_2023}
Morishita, T., \& Stiavelli, M. 2023, The Astrophysical Journal Letters, 946,
  L35, \dodoi{10.3847/2041-8213/acbf50}

\bibitem[{{Ngol{\`e} Mboula, F. M.} {et~al.}(2015){Ngol{\`e} Mboula, F. M.},
  {Starck, J.-L.}, {Ronayette, S.}, {Okumura, K.}, \& {Amiaux,
  J.}}]{Mboula+2015}
{Ngol{\`e} Mboula, F. M.}, {Starck, J.-L.}, {Ronayette, S.}, {Okumura, K.}, \&
  {Amiaux, J.} 2015, A\&A, 575

\bibitem[{{Nie} {et~al.}(2021{\natexlab{a}}){Nie}, {Li}, {Peterson}, \&
  {Wei}}]{2021MNRAS.503.4436N}
{Nie}, L., {Li}, G., {Peterson}, J.~R., \& {Wei}, C. 2021{\natexlab{a}},
  \mnras, 503, 4436, \dodoi{10.1093/mnras/stab733}

\bibitem[{{Nie} {et~al.}(2021{\natexlab{b}}){Nie}, {Li}, {Zhang}, {Fan}, \&
  {Peterson}}]{2021MNRAS.508.3785N}
{Nie}, L., {Li}, G., {Zhang}, J., {Fan}, Z., \& {Peterson}, J.~R.
  2021{\natexlab{b}}, \mnras, 508, 3785, \dodoi{10.1093/mnras/stab2824}

\bibitem[{{Nie} {et~al.}(2024){Nie}, {Shan}, {Li}, {Wang}, {Cheng}, {Tao},
  {Cui}, {Xie}, {Liu}, \& {Zhang}}]{2024AJ....167...58N}
{Nie}, L., {Shan}, H., {Li}, G., {et~al.} 2024, \aj, 167, 58,
  \dodoi{10.3847/1538-3881/ad14f7}

\bibitem[{{Ono} {et~al.}(2023){Ono}, {Harikane}, {Ouchi}, {Yajima}, {Abe},
  {Isobe}, {Shibuya}, {Wise}, {Zhang}, {Nakajima}, \&
  {Umeda}}]{2023ApJ...951...72O}
{Ono}, Y., {Harikane}, Y., {Ouchi}, M., {et~al.} 2023, \apj, 951, 72,
  \dodoi{10.3847/1538-4357/acd44a}

\bibitem[{Pearson(1901)}]{pearson1901liii}
Pearson, K. 1901, The London, Edinburgh, and Dublin Philosophical Magazine and
  Journal of Science, 2, 559

\bibitem[{{Perrin} {et~al.}(2015){Perrin}, {Long}, {Sivaramakrishnan},
  {Lajoie}, {Elliot}, {Pueyo}, \& {Albert}}]{2015ascl.soft04007P}
{Perrin}, M.~D., {Long}, J., {Sivaramakrishnan}, A., {et~al.} 2015, {WebbPSF:
  James Webb Space Telescope PSF Simulation Tool}, Astrophysics Source Code
  Library, record ascl:1504.007.
\newblock \doeprint{1504.007}

\bibitem[{{Perrin} {et~al.}(2014){Perrin}, {Sivaramakrishnan}, {Lajoie},
  {Elliott}, {Pueyo}, {Ravindranath}, \& {Albert}}]{2014SPIE.9143E..3XP}
{Perrin}, M.~D., {Sivaramakrishnan}, A., {Lajoie}, C.-P., {et~al.} 2014, in
  Society of Photo-Optical Instrumentation Engineers (SPIE) Conference Series,
  Vol. 9143, Space Telescopes and Instrumentation 2014: Optical, Infrared, and
  Millimeter Wave, ed. J.~{Oschmann}, Jacobus~M., M.~{Clampin}, G.~G. {Fazio},
  \& H.~A. {MacEwen}, 91433X, \dodoi{10.1117/12.2056689}

\bibitem[{{Perrin} {et~al.}(2012){Perrin}, {Soummer}, {Elliott}, {Lallo}, \&
  {Sivaramakrishnan}}]{2012SPIE.8442E..3DP}
{Perrin}, M.~D., {Soummer}, R., {Elliott}, E.~M., {Lallo}, M.~D., \&
  {Sivaramakrishnan}, A. 2012, in Society of Photo-Optical Instrumentation
  Engineers (SPIE) Conference Series, Vol. 8442, Space Telescopes and
  Instrumentation 2012: Optical, Infrared, and Millimeter Wave, ed. M.~C.
  {Clampin}, G.~G. {Fazio}, H.~A. {MacEwen}, \& J.~{Oschmann}, Jacobus~M.,
  84423D, \dodoi{10.1117/12.925230}

\bibitem[{{Pontoppidan} {et~al.}(2022){Pontoppidan}, {Barrientes}, {Blome},
  {Braun}, {Brown}, {Carruthers}, {Coe}, {DePasquale}, {Espinoza}, {Marin},
  {Gordon}, {Henry}, {Hustak}, {James}, {Jenkins}, {Koekemoer}, {LaMassa},
  {Law}, {Lockwood}, {Moro-Martin}, {Mullally}, {Pagan}, {Player}, {Proffitt},
  {Pulliam}, {Ramsay}, {Ravindranath}, {Reid}, {Robberto}, {Sabbi}, {Ubeda},
  {Balogh}, {Flanagan}, {Gardner}, {Hasan}, {Meinke}, \&
  {Nota}}]{2022ApJ...936L..14P}
{Pontoppidan}, K.~M., {Barrientes}, J., {Blome}, C., {et~al.} 2022, \apjl, 936,
  L14, \dodoi{10.3847/2041-8213/ac8a4e}

\bibitem[{Rowe {et~al.}(2011)Rowe, Hirata, \& Rhodes}]{Rowe+2011}
Rowe, B., Hirata, C., \& Rhodes, J. 2011, The Astrophysical Journal, 741, 46,
  \dodoi{10.1088/0004-637X/741/1/46}

\bibitem[{Starck {et~al.}(2002)Starck, Pantin, \& Murtagh}]{Starck+2002}
Starck, J.~L., Pantin, E., \& Murtagh, F. 2002, Publications of the
  Astronomical Society of the Pacific, 114, 1051, \dodoi{10.1086/342606}

\bibitem[{Symons {et~al.}(2021)Symons, Zemcov, Bock, Cheng, Crill, Hirata, \&
  Venuto}]{Symons+2021}
Symons, T., Zemcov, M., Bock, J., {et~al.} 2021, The Astrophysical Journal
  Supplement Series, 252, 24, \dodoi{10.3847/1538-4365/abcaa5}

\bibitem[{Takeda {et~al.}(2006)Takeda, Farsiu, Christou, \&
  Milanfar}]{Takeda+2006}
Takeda, H., Farsiu, S., Christou, J., \& Milanfar, P. 2006, in The Maui
  Economic Development Board, ed. S.~Ryan, Vol. id.E27, Advanced Maui Optical
  and Space Surveillance (AMOS) Technologies Conference.
\newblock \url{https://api.semanticscholar.org/CorpusID:5210235}

\bibitem[{Wang {et~al.}(2022)Wang, Li, \& Kang}]{Wang_2022}
Wang, L., Li, G., \& Kang, X. 2022, Monthly Notices of the Royal Astronomical
  Society, 517, 787, \dodoi{10.1093/mnras/stac2664}

\bibitem[{Wang \& Li(2017)}]{Wang_2017}
Wang, L., \& Li, G.-L. 2017, Research in Astronomy and Astrophysics, 17, 100,
  \dodoi{10.1088/1674-4527/17/10/100}

\bibitem[{Wang {et~al.}(2024)Wang, Shan, Nie, Liu, Yan, Li, Cheng, Xie, Qu,
  Zheng, \& Kang}]{Wang_2024}
Wang, L., Shan, H., Nie, L., {et~al.} 2024, Research in Astronomy and
  Astrophysics, 24, 045009, \dodoi{10.1088/1674-4527/ad2edf}

\end{thebibliography}

\end{document}